\documentclass[useAMS,usenatbib]{mn2e}

\usepackage{graphicx}
\usepackage{epsfig,floatflt}

\newcommand \approxgt{\,\raise2pt \hbox{$>$}\kern-8pt\lower2.pt\hbox{$\sim$}\,}
\newcommand \approxlt{\,\raise2pt \hbox{$<$}\kern-8pt\lower2.pt\hbox{$\sim$}\,}

\title[Cepheid investigations using  the {\it Kepler} space telescope]{Cepheid
investigations using the {\it Kepler} space telescope}
\author[R. Szab\'o, L. Szabados, C.C. Ngeow, et~al.]
{R. Szab\'o$^{1}$\thanks{E-mail: rszabo@konkoly.hu},
L. Szabados$^{1}$, C.-C. Ngeow$^{2}$, R. Smolec$^{3}$, A. Derekas$^{1,4}$\thanks{Magyary Zolt\'an Postdoctoral Research Fellow}, \newauthor
P. Moskalik$^{5}$, J. Nuspl$^{1}$, H. Lehmann$^{6}$, G. F\H ur\'esz$^{7}$, J. Molenda-\.Zakowicz$^{8}$,\newauthor
S.~T. Bryson$^{9}$, A.~A. Henden$^{10}$, D.~W. Kurtz$^{11}$,  D. Stello$^{12}$, J.~M. Nemec$^{13}$, \newauthor
J.~M. Benk\H o$^{1}$, L. Berdnikov$^{14,15}$, H. Bruntt$^{16}$, N.~R. Evans$^{17}$, N.~A. Gorynya$^{18}$, \newauthor
E.~N. Pastukhova$^{18}$, R.~J. Simcoe$^{14}$, J.~E. Grindlay$^{19}$, E.~J. Los$^{19}$, A. Doane$^{19}$, \newauthor
S.~G. Laycock$^{19}$, D.~J. Mink$^{17}$, G. Champine$^{19}$, A. Sliski$^{19}$, G. Handler$^{3}$, \newauthor
L.~L. Kiss$^{1}$, Z. Koll\'ath$^{1}$, J. Kov\'acs$^{20}$, J.~Christensen-Dalsgaard$^{21}$, H.~Kjeldsen$^{21}$, \newauthor
C. Allen$^{22}$, S.~E. Thompson$^{23}$, J. Van Cleve$^{23}$
\\
$^{1}$Konkoly Observatory of the Hungarian Academy of Sciences, Konkoly Thege Mikl\'os \'ut 15-17, H-1121 Budapest,  Hungary\\
$^{2}$Graduate Institute of Astronomy, National Central University, Jhongli City, Taoyuan County 32001, Taiwan\\
$^{3}$Institut f\"ur Astronomie, Universit\"at Wien, T\"urkenschanzstrasse 17, A-1180 Wien, Austria\\
$^{4}$Department of Astronomy, E\"otv\"os University, Budapest, Hungary\\
$^{5}$Copernicus Astronomical Center, ul. Bartycka 18, 00-716, Warsaw, Poland\\
$^{6}$Th\"uringer Landessternwarte Tautenburg, Karl-Schwarzschild-Observatorium, 07778 Tautenburg, Germany\\
$^{7}$Harvard-Smithsonian Center for Astrophysics, MS 20, 60 Garden Street, Cambridge, MA 02138, USA\\
$^{8}$Astronomical Institute, University of Wroc\l{}aw, ul. Kopernika 11, 51-622, Wroc\l{}aw, Poland\\
$^{9}$NASA Ames Research Center, Moffett Field, CA 94035, USA\\
$^{10}$American Association of Variable Star Observers, Cambridge, MA 02138, USA\\
$^{11}$Jeremiah Horrocks Institute of Astrophysics, University of Central Lancashire, Preston PR1 2HE, UK\\
$^{12}$Sydney Institute for Astronomy, School of Physics, The University of Sydney, Sydney, 2006 NSW, Australia\\
$^{13}$Department of Physics \& Astronomy, Camosun College, Victoria, British Columbia, V8P 5J2, Canada\\
$^{14}$Sternberg Astronomical Institute, Moscow University, Moscow, Russia\\
$^{15}$Isaac Newton Institute of Chile, Moscow Branch, Universitetskij Pr. 13, Moscow 119992, Russia\\
$^{16}$LESIA, UMR 8109, Observatoire de Paris, 92195 Meudon Cedex, France\\
$^{17}$Smithsonian Astrophysical Observatory, MS 4, 60 Garden Street, Cambridge, MA 02138, USA\\
$^{18}$Institute of Astronomy, Russian Academy of Sciences, 48 Pyatnitskaya Street, Moscow, 109017, Russia\\
$^{19}$Harvard College Observatory, 60 Garden Street, Cambridge, MA 02138, USA\\
$^{20}$Gothard Astrophysical Observatory, H-9707, Szombathely, Hungary\\
$^{21}$Department of Physics and Astronomy, Aarhus University, DK-8000 Aarhus C, Denmark\\
$^{22}$Orbital Sciences Corporation/NASA Ames Research Center, Moffett Field, CA 94035\\
$^{23}$SETI Institute/NASA Ames Research Center, Moffett Field, CA 94035}
\begin{document}

\date{Accepted 2011 January 12, Received 2010 December 31}


\maketitle

\label{firstpage}

\begin{abstract}
We report results of initial work done on selected candidate Cepheids 
to be observed with the {\it Kepler} space telescope. Prior to the launch 40  
candidates were selected from previous surveys and databases. The analysis of the 
first 322 days of {\it Kepler} photometry, and recent ground-based follow-up multicolour 
photometry and spectroscopy allowed us to confirm that one of these stars, V1154\,Cyg 
(KIC\,7548061), is indeed a 4.9-d Cepheid. Using the phase lag method we show that 
this star pulsates in the fundamental mode. New radial velocity data are consistent 
with previous measurements, suggesting that a long-period binary component is unlikely.
No evidence is seen in the ultra-precise, nearly uninterrupted {\it Kepler} photometry 
for nonradial or stochastically excited modes at the micromagnitude level.
The other candidates are not Cepheids but an interesting mix of possible spotted stars, 
eclipsing systems and flare stars. 
\end{abstract}

\vspace*{2mm}

\section{Introduction}

The {\it Kepler}\footnote{http://kepler.nasa.gov} space mission is designed 
to detect Earth-like planets around solar-type stars with the transit method \citep{BKB10}
by monitoring continuously over 150\,000 stars with an unprecedented 
photometric precision. The lifetime of at least 3.5 years and the 
quasi-continuous observations make {\it Kepler} an ideal tool to measure stellar photometric 
variability with a precision that is unachievable from the ground (see, e.g.
\citealt{gil10a}).

Ground-based photometric observations of Cepheids usually consist of a few (typically 
one or two) observations per night. Until now it has not been possible to 
adequately cover many consecutive pulsational cycles with ground-based 
photometry. Ground-based studies of several types of variable stars -- such 
as $\delta$\,Sct stars, DOV, DBV and DAV white dwarfs, sdBV stars and roAp stars -- 
have benefited from multisite photometric observations. But because of their
longer period, Cepheids have not been the targets of such concerted efforts,  
with the notable exception of V473\,Lyr \citep{BSA86}.

Space-based Cepheid observations were conducted earlier with the star tracker of 
the {\sc WIRE} satellite and the Solar Mass Ejection Imager (SMEI) instrument on 
board the Coriolis satellite \citep{bes08, ss08, berd10}. The lengths of 
these data sets ($\sim 1000-1600$\,d) are comparable to the nominal lifetime of the 
{\it Kepler} Mission. The WIRE and SMEI data of Polaris ($\alpha$\,UMi, 
$V=2.005$) confirmed that the pulsation amplitude of Polaris has been increasing again,
after a long period of decrease. {\it Kepler} is capable of delivering 
comparably accurate photometric observations for the much fainter Cepheid, 
V1154\,Cyg ($V=9.19$).


The Kepler Asteroseismic Science Consortium 
{\sc kasc}\footnote{http://astro.phys.au.dk/KASC/} was set up to exploit the 
{\it Kepler} data for studying stellar pulsations. {\sc kasc} working group 
number 7 (WG7) is dedicated to the investigation of Cepheids. In compliance with 
one of the original {\sc kasc} proposals submitted before the launch of the space 
telescope, we searched for Cepheids among a list of 40 targets, including the only 
previously known Cepheid V1154\,Cyg (KIC\,7548061) in the field. In this paper we 
describe the first results from {\it Kepler} observations of a Cepheid and 
a couple of not confirmed Cepheids, complemented by extended ground-based 
follow-up observations.

\begin{figure}
\includegraphics[height=82mm, angle=270]{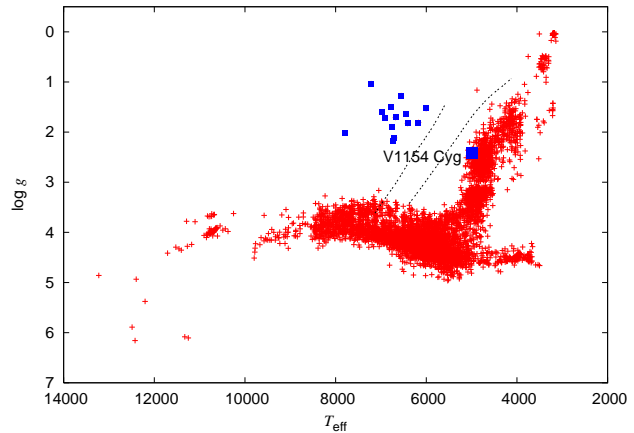}
\caption{Selection of Cepheid candidates based on Kepler Input Catalog effective 
temperature and $\log g$ values (small squares). The plot contains 5987 {\sc kasc} 
targets (red plus signs); the selected Cepheid candidates are shown by full (blue) 
squares. V1154\,Cyg, the only confirmed Cepheids is denoted by a large (blue) square. 
The linear Cepheid instability strip is denoted by dashed lines.}
\label{KIC}
\end{figure}

\begin{table*}
\caption{Main properties of the observed {\it Kepler} Cepheid and Cepheid 
candidates. The uncertainty of the period is given in parentheses and refers to 
the last digit(s) of the period. Q3.1 means the first month of Q3. See text for 
the definition of the contamination index.}
\label{tab1}
\begin{center}
\begin{tabular}{rlrrrlcl}
\hline
\multicolumn{1}{c}{KIC ID} & \multicolumn{1}{c}{Other name} & 
\multicolumn{1}{c}{$\alpha$} & \multicolumn{1}{c}{$\delta$} &  
\multicolumn{1}{c}{$K_p$}  & \multicolumn{1}{c}{Period}   & 
\multicolumn{1}{c}{Contam.}
 &  \multicolumn{1}{c}{Runs} \\
         &           & \multicolumn{1}{c} {J2000} &  \multicolumn{1}{c}{J2000}    
&  \multicolumn{1}{c}{mag}  & \multicolumn{1}{c}{days}  &\multicolumn{1}{c}{Index} 
&\\ 
\hline
{\bf 7548061} & {\bf V1154 Cyg} & {\bf 19 48 15.45} &   {\bf  +43 07 36.77} & {\bf
8.771} & {\bf  4.925454(1)} & 0.036 & {\bf LC: Q01234 SC: Q1}\\ \hline
2968811 & ASAS 190148+3807.0  & 19 01 48.21 &  +38 07 01.99 & 13.469& 14.8854(12) 
&
0.047 & LC: Q01234\\
6437385 & ASAS 192000+4149.1  & 19 20 00.12 &  +41 49 07.46 & 11.539&  13.6047(7) 
&
0.075 & LC: Q01234\\
8022670 & V2279 Cyg & 19 18 54.46 &    +43 49 25.82 &12.471 & 4.12564(5) & 0.189  
& LC:
Q1234\\
12406908 & ROTSE1  & 19 23 44.99 &+51 16 11.75 & 12.354 &
13.3503(45)&
0.015& LC: Q01 SC: Q3.1\\
& J192344.95+511611.8   &  & &  & && \\
\hline
\end{tabular}
\end{center}
\end{table*}


\section{Target selection} \label{tsel}

Being supergiants, Cepheids are rare and the number of Cepheids in the {\it Kepler} 
fixed 105 square degree field of view (FOV) is expected to be low. 
Due to telemetry bandwidth constraints, {\it Kepler} cannot observe all
stars in the field.  The first 10 months of operation (observing quarters
Q0--Q4) were therefore dedicated to a survey phase to find the best
candidates, which would then be observed during the rest of the mission.

We used two different approaches to find Cepheid candidates for the survey
phase. First, all known, possible or suspected variable stars in the relevant 
pulsational period range were selected from all available databases containing
variability information, such as GCVS \citep{gcvs}, ASAS North \citep{asas09}, 
ROTSE \citep{akerlof00} and HAT catalogs \citep{hat04}. Light curve shapes and the 
$\log P$ vs. $J-H$ diagram \citep{pm04} were also utilized for further selection. 
Finally, stars with close (bright) companions 
were excluded. This procedure resulted in a list of 26 stars. We note that 
these catalogs are not complete in coverage of the field, nor in
the relevant magnitude range. 

As a second approach, we searched for stars lying inside or close to the
Cepheid instability strip based on the Kepler Input Catalog
(KIC)\footnote{http://archive.stsci.edu/kepler} effective temperature  and 
$\log g$ values. Stars fainter than $Kp = 16.0$\,mag were excluded, where $Kp$ 
denotes the {\it Kepler} magnitude system (see Section\,\ref{obs}. for more 
details on the {\it Kepler} magnitudes). Candidates with a contamination index 
(CI) larger than 0.5 were also omitted, where ${\rm CI} = 0$ means that all the 
flux in the aperture comes from the target and in case of ${\rm CI} = 1$ the 
flux entirely comes from surrounding sources. This resulted in 14 additional 
targets as shown in Fig.\,\ref{KIC}. The linear Cepheid instability strip, denoted 
by dashed lines in the figure, was calculated with the Florida-Budapest code 
\citep{sbb07}.

We note that the process followed to derive the KIC parameters was optimized 
to find main-sequence stars with high probability and to distinguish
them from cool giants \citep{bbk10}. The precision of the parameters is 
therefore limited for our purposes. This is illustrated in Fig.\,\ref{KIC} by the 
fact the KIC parameters of V1154\,Cyg (the only previously known Cepheid in the 
field) put it outside the computed instability strip by several hundred Kelvins.

Consequently, we proposed to observe the above mentioned 40 stars by {\it Kepler} 
during the `survey period'. 


\section{{\it Kepler} observations}\label{obs}

\begin{figure}
\includegraphics[height=78mm, angle=0]{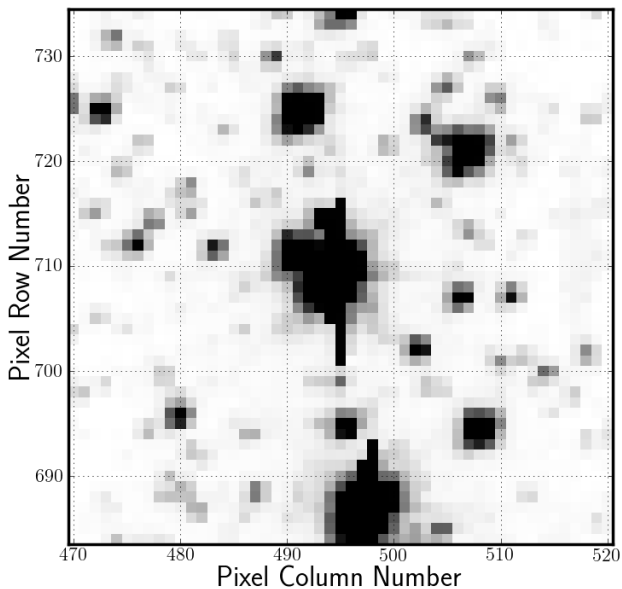}
\caption{Vicinity of  V1154\,Cyg from one of the Full Frame Images. The star
is slightly saturated as indicated by its brightness and the elongated shape.}
\label{FFI}
\end{figure}

{\it Kepler} was launched on 2009 March 6 into a 372-d solar orbit, and is observing a 
105 square degree area of the sky in between the constellations of Cygnus and Lyra \citep{koch10}.
After a 9.7\,d commissioning phase (Q0), the regular observations started on 2009 May 12. 
In order to ensure optimal solar illumination of the solar panels, a 90\,degree 
roll of the telescope is performed at the end of each quarter of its solar orbit. 
The first quarter lasted only for 33.5\,d (Q1), while subsequent quarters are all three months
long. In each of the four quarters annually the {\it Kepler} targets fall on different CCDs.

The {\it Kepler} magnitude system ($Kp$) refers to the wide passband ($430-900$\,nm)
transmission of the telescope and detector system. Note that the {\it Kepler} magnitudes 
($Kp$) were derived before the mission launch and are only approximate values. Currently 
{\it Kepler} processing does not provide calibrated {\it Kepler} magnitudes $Kp$ \citep{kk10b}.
Both long cadence (LC, 29.4\,min, \citealt{jen10b}), and short cadence (SC, 58.9\,s, 
\citealt{gil10b}) observations are based on the same $6$-s integrations which are summed 
to form the LC and SC data onboard. In this work we use BJD-corrected, raw LC data \citep{jen10a} 
spanning from Q0 to Q4, i.e. 321.7\,d of quasi-continuous observations. Some of our targets  
(V1154\,Cyg among them) were observed in SC mode as well. We exploit this opportunity to 
compare LC and SC characteristics and investigate the frequency spectrum to a much higher
Nyquist frequency (733.4\,d$^{-1}$ vs. 24.5\,d$^{-1}$). 

The saturation limit is between
$Kp \simeq 11-12$\,mag depending on the particular chip the star is
observed; brighter than this, accurate photometry can be performed 
up to $Kp \simeq 7$\,mag with judiciously designed apertures \citep{szr10}. 
Since V1154\,Cyg is much brighter than the saturation limit, it required 
special treatment, as can be seen in Fig.\,\ref{FFI}, which shows a 50 
pixel box centered on V1154\,Cyg. The plot was made using 
{\sc keplerffi}\footnote{http://keplergo.arc.nasa.gov/ContributedSoftwareKeplerFFI.shtml}
written by M.~Still. 

\begin{figure}
\includegraphics[width=58mm, angle=270]{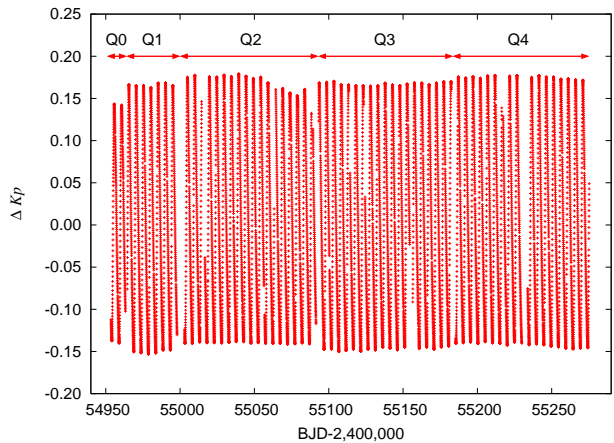}
\caption{Raw {\it Kepler} light curve of V1154\,Cyg.  
The varying amplitude is instrumental in origin; see text for more details.}
\label{v1154lc}
\end{figure}

\begin{figure}
\includegraphics[width=58mm, angle=270]{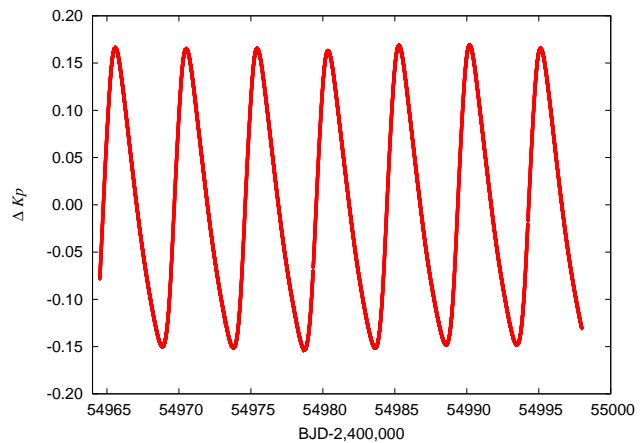}
\caption{Q1 SC {\it Kepler} light curve of V1154\,Cyg. 
It is almost indistinguishable from LC data taken during 
the same time interval. Note the essentially uninterrupted sampling.}
\label{v1154sc}
\end{figure}

To illustrate some of the common characteristics of the data we show in 
Fig.\,\ref{v1154lc} the raw {\it Kepler} light curve of V1154\,Cyg after
normalizing the raw flux counts and converting the fluxes to the magnitude
scale. The small gaps in the light curve are due to unplanned safe mode and 
loss-of-fine-point events, as well as regular data downlink periods. This LC 
data set spanning Q0--Q4 data contains 14485 points. Fig.\,~\ref{v1154sc} shows 
a 33.5\,d segment (Q1) where SC data are available for V1154\,Cyg containing 
49032 data points. 

The varying amplitude seen in Fig.\,~\ref{v1154lc} is of 
instrumental origin. It is a result of the small drift of the telescope, 
coupled with different pixel sensitivities. In addition, different aperture 
masks are assigned to the targets quarterly which result in small changes in 
the measured flux. The most notable amplitude change is seen towards the end 
of Q2, which was noted to be due to flux flowing outside of the optimal 
aperture (bleeding), affecting the measured brightness. Fortunately, in Q2 a 
larger mask was also downloaded besides the standard optimal aperture assigned 
to this star which allowed us to investigate the variation of flux outside 
the optimal aperture. This confirmed that the total flux was indeed captured 
within the larger aperture, and hence that the star shows no intrinsic amplitude 
variation within the current accuracy of the data. Without further information 
we cannot choose between a possible slight amplitude variation and instrumental 
effects in other quarters. 

\section{Target classification} \label{vetting}

After the release of the first {\it Kepler} light curves to the {\sc kasc}  
it turned out that only a few stars showed Cepheid like variability. 
The properties of these {\it Kepler} Cepheid candidates are listed in 
Table\,\ref{tab1}. Periods were determined for the {\it Kepler} LC 
light curves using the {\sc period04} program \citep{lb05}.

Apart from these candidates a number of other variable stars were included 
in the initial sample of 40 stars. Based on their light curves and the automatic 
classification \citep{BDDR10} we classify these as eclipsing binaries (8 stars), 
ellipsoidal variables (6), delta Scuti stars (3), an SPB star (1), long period 
variables (7) and stars with no obvious variations (7). While this sample is a 
rich source of flaring, spotted, granulated stars, the detailed investigation of 
these non-Cepheids is out of scope of this paper. In addition, we inspected all 
6300 {\sc kasc} light curves, but found no Cepheids among them.

Three originally proposed stars were
not observed due to technical reasons (position on the CCD chip, 
brightness/faintness, contamination or pixel number constraints). 
The large variety of non-Cepheids in our sample indicates that
the spatial resolution and photometric accuracy of 
ASAS was not adequate to select such relatively faint Cepheids.  It further
shows that the Kepler Input Catalog is clearly not optimal for our purpose.


\section{Ground-based follow up}\label{ground}
To further confirm or disprove the Cepheid nature of our candidates in 
Table\,\ref{tab1} we employed ground-based multicolour photometry and 
in some cases spectroscopy. In the case of V1154\,Cyg regular radial velocity
observations and multicolour photometry were scheduled, which helped us 
to gain more information on the pulsation. In the following we describe these 
ground-based follow-up observations and the subsequent scrutinizing
classification process on a star-by-star basis. 

\subsection{Multicolour photometry}\label{mulcol}

Ground-based multicolour observations provide additional information,
therefore complement the space-born photometry taken in `white light'. 
We used the following telescopes to gather $BVR_cI_c$ Johnson-Cousins 
magnitudes:

\begin{enumerate}
\item{Lulin One-meter Telescope (LOT) at Lulin Observatory 
(Taiwan)\footnote{http://www.lulin.ncu.edu.tw/english/index.htm}: 
1-m Cassegrain-telescope with PI1300B CCD;}

\item{SLT at Lulin Observatory (Taiwan): 0.4-m RC-tele\-scope with Apogee U9000
CCD;}

\item{Tenagra telescope at Tenagra II Observatory\footnote{http://www.tenagraobservatories.com} 
(USA): robotic 0.8-m RC-telescope with SITe CCD;}

\item{Sonoita Research Observatory (SRO, USA)\footnote{http://www.sonoitaobservatories.org},  
0.35-m robotic telescope with a SBIG STL-1001E CCD.}
\end{enumerate}

Observations were performed from September 2009 to August 2010. 
All of the imaging data were reduced in a standard way 
(bias-subtracted, dark-subtracted and flat-fielded) using 
{\sc iraf}\footnote{{\sc iraf} is distributed by the National Optical Astronomy Observatories, 
which are operated by the Association of Universities for Research in Astronomy, 
Inc., under cooperative agreement with the National Science Foundation.}. 
Instrumental magnitudes for the stars in the images were measured using 
{\tt SExtractor} \citep{BA96} with aperture photometry. For the SRO 
data a separate photometric pipeline based on {\sc daophot} was used, 
which included analysis of an ensemble of comparison stars. 
The instrumental magnitudes were transformed to the standard magnitudes using 
Landolt standards \citep{Land09}. Times of observation were converted to  
heliocentric Julian day.

\begin{figure*}
\includegraphics[width=165mm, angle=0]{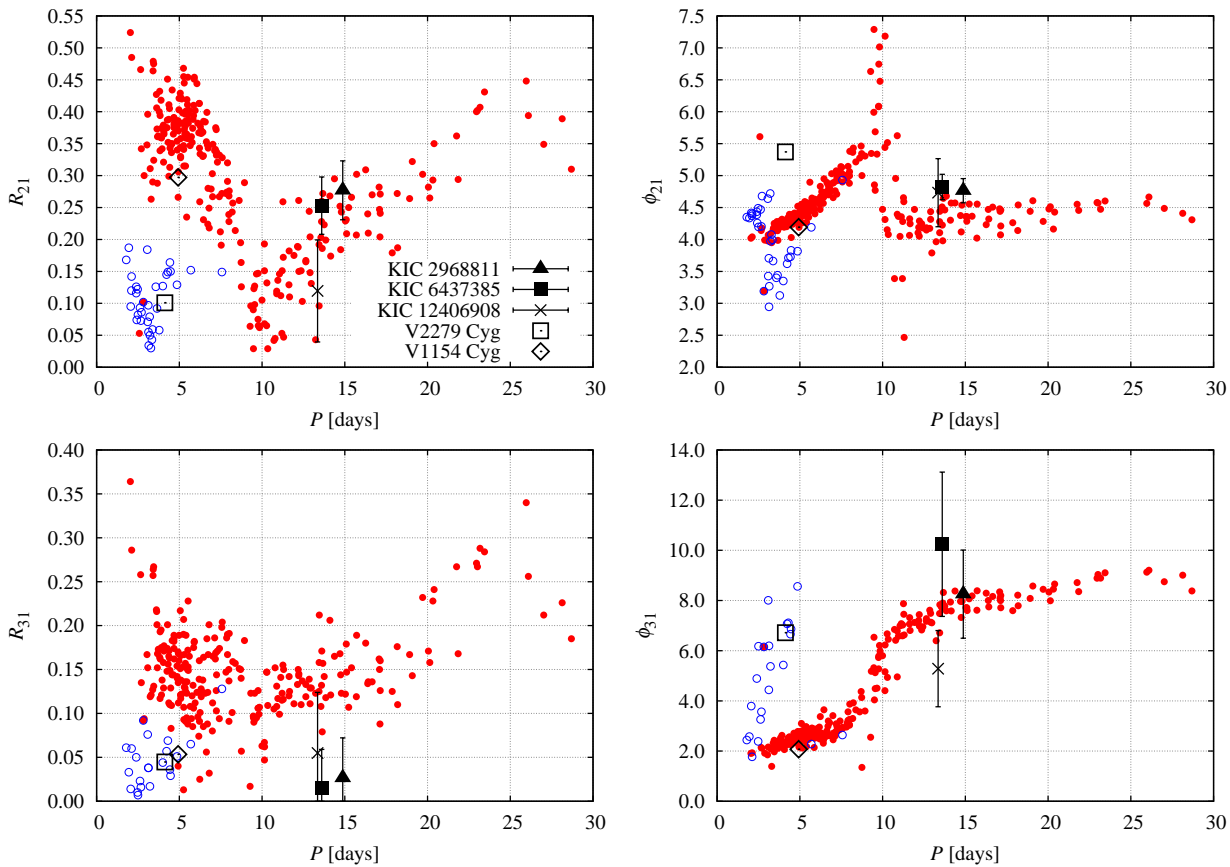}
\caption{Fundamental mode (red filled) and first overtone (blue open symbols) 
Fourier parameter progressions in Johnson $V$ passband for Galactic Cepheid sample. 
V1154\,Cyg and other {\it Kepler} Cepheid candidates are also shown based on
their new ground-based Johnson $V$ photometry. Error bars for V1154\,Cyg 
and V2279\,Cyg are comparable to, or smaller than the symbol size, so these were omitted 
from the plot.The Galactic Cepheid light curve parameters were compiled from 
\citet{fp1}, \citet{fp2}, \citet{fp3}, \citet{fp4}, \citet{fp5}, \citet{fp6},
\citet{fp7}.}
\label{fps}
\end{figure*}

We gathered between 80 and 120 frames per target for each passband, 
excluding KIC\,12406908 which was observed only at TNG and hence had 
fewer images taken. We have found a small offset 
between SRO and other $B$ data in the case of V1154\,Cyg. Fortunately, 
data were taken in TNG and SRO very close in time (less than 1 minute) twice, 
allowing us to correct the SRO data by the shift measured in these epochs: 
$B_{\rm SRO}-B_{\rm TNG}=0.054$. All the (corrected) photometric measurements 
are available online, while Table\,\ref{onl1} shows the layout of the data.

\begin{table}
\caption{Ground-based multicolour photometry of V1154\,Cyg and {\it Kepler} 
Cepheid candidates. The entire table is available only electronically. }
\label{onl1}
\begin{tabular}{cccccc}
\hline
{KIC ID} & HJD & Mag. & Err. &  Filt.  & Obs.   \\ \hline
02968811  &  2455103.7700  &	14.857 & 0.013 &  B & SRO\\
02968811  &  2455103.7809  &	14.823 & 0.013 &  B & SRO\\
02968811  &  2455104.7544  &	14.797 & 0.011 &  B & SRO\\
...  & ... &...	 & ... & ...  & ...\\
\hline
\end{tabular}
\end{table}

We decomposed the multicolour light curves to derive the widely used Fourier 
parameters $R_{\rm i1}$ and $\phi_{\rm i1}$, as defined by \citet{sile81}, 
which characterize the light curve shape. 
The Johnson $V$ results are plotted in Fig.\,\ref{fps} along with the Fourier parameters
of fundamental mode (red filled points) and first overtone (blue open circles) 
Galactic Cepheids as a function of pulsational period. As all
Galactic Cepheids follow the main Fourier progressions in Fig.\,\ref{fps}, 
a star that lies outside the overall pattern is unlikely to be Cepheid. 
We discuss our findings for individual candidates below. 

\subsection{Spectroscopy}

\begin{figure*}
$\begin{array}{cc}
\includegraphics[angle=270,scale=0.3]{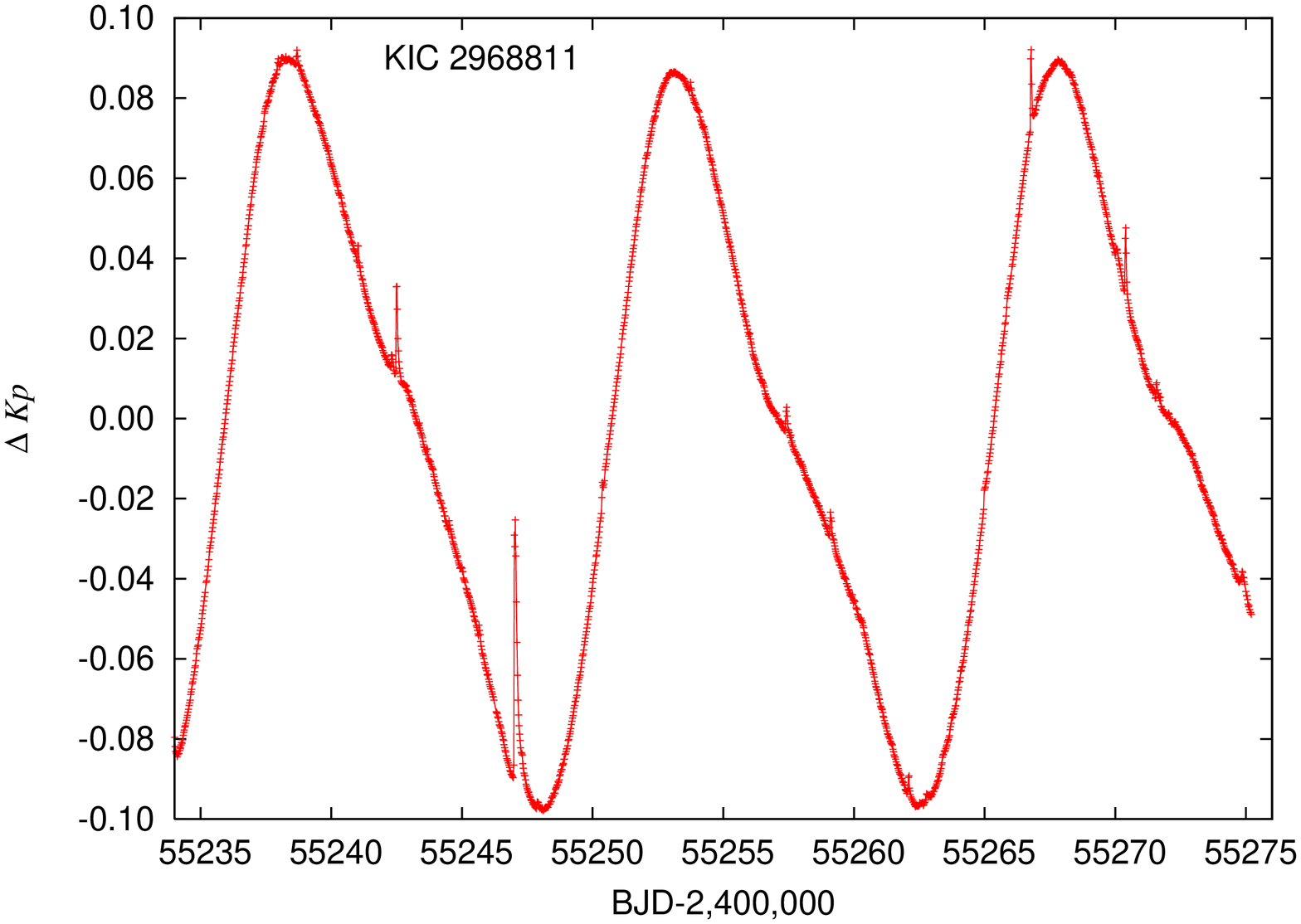} &
\includegraphics[angle=270,scale=0.3]{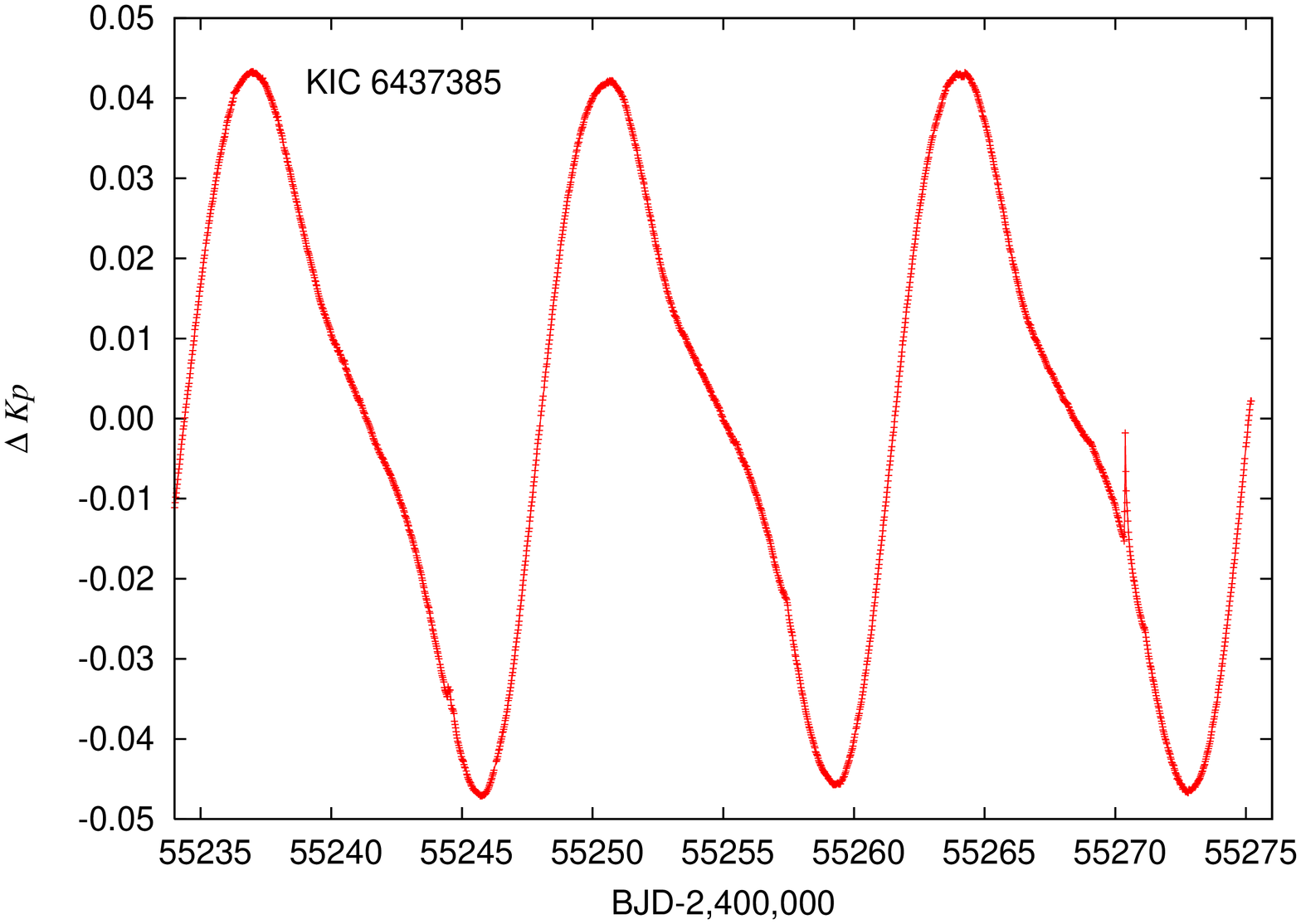} \\
\includegraphics[angle=270,scale=0.3]{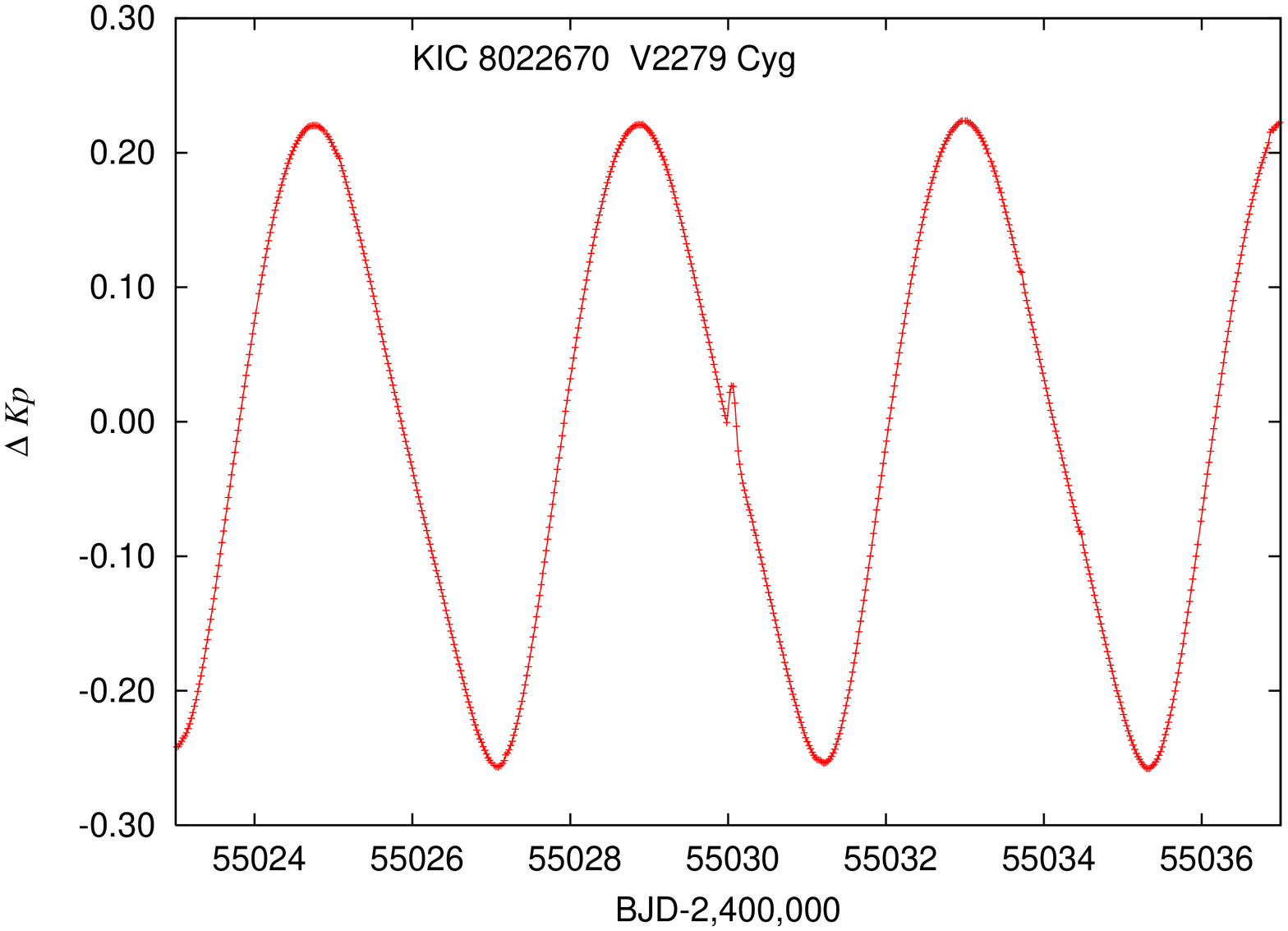} &
\includegraphics[angle=270,scale=0.3]{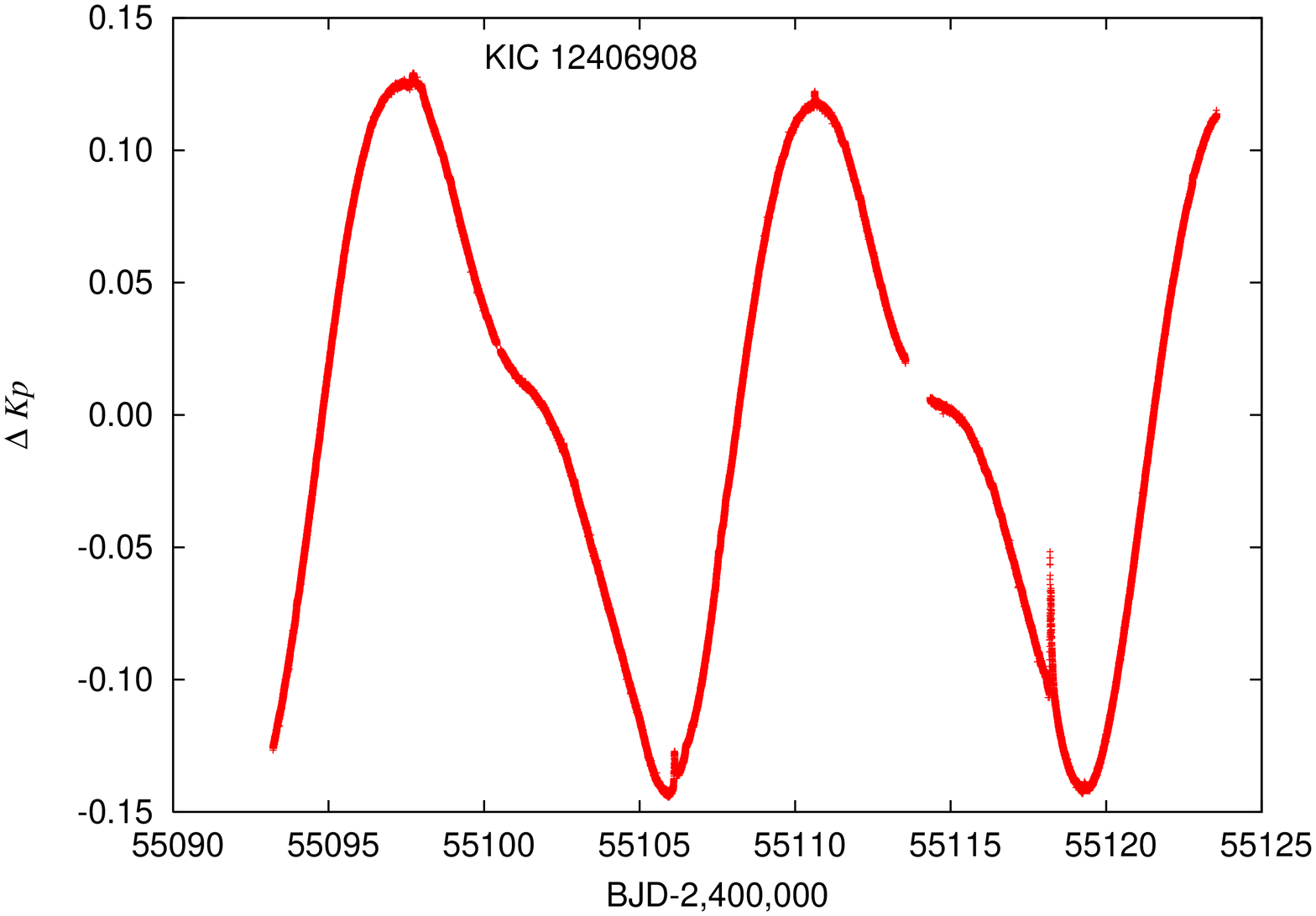} \\
\end{array}$
\caption{Representative parts of {\it Kepler} photometry of four Cepheid 
candidates (excluding the known Cepheid V1154\,Cyg). LC light curves are 
plotted, except for KIC\,12406908 where one month of SC data is plotted. 
The `shoulders' on the descending branches indicate spotted stars and the
flares are typical for active stars. These light curves indicate that
KIC\,8022670 (V2279~Cyg) is the most promising Cepheid candidate of 
the four. \label{kepphot}}
\end{figure*}

Spectroscopic observations were conducted to measure radial velocity and derive 
stellar parameters. We obtained spectra of V1154\,Cyg with the Coude-Echelle 
spectrograph attached to the 2-m telescope of the Th\"uringer Landessternwarte 
(TLS) Tautenburg. Spectra cover the region $470 - 740$\,nm, the spectral
resolution is $R = 33\,000$, the exposure time was 30\,min per spectrum. Spectra 
were reduced using standard {\sc midas} packages. The reduction included the 
removal of cosmic rays, bias and stray light subtraction, flat fielding, optimum 
order extraction, wavelength calibration using a Th-Ar lamp, normalization to the 
local continuum, and merging of the orders. Small instrumental shifts were 
corrected by an additional calibration using a larger number of 
telluric O$_2$ lines. In addition, one spectrum of V1154\,Cyg was taken on 2007 
June 15 at the {\it M. G. Fracastoro} station (Serra La Nave, Mt. Etna) of the 
{\it INAF - Osservatorio Astrofisico di Catania} (INAF-OACt). We used the 91-cm 
telescope and {\sc fresco}, the fiber-fed {\sc  reosc} echelle spectrograph which 
allowed us to obtain spectra in the range of $4300-6800$\,\AA\ with a resolution 
$R=21\,000$.

For V2279\,Cyg we made a high-resolution spectrum in a single-shot 1680~s exposure 
with the 1.5-m tele\-scope at the Fred Lawrence Whipple Observatory (Mt. Hopkins, 
Arizona) using the Tillinghast Reflection Echelle Spectrograph (TRES) on 2010 
September 24. The spectrum is cross-dispersed with a wavelength range of $3860-
9100$\,\AA\ over 51 spectral orders.

\subsection{Remarks on individual candidates}

In the following we summarize the {\it Kepler} light curve characteristics, 
the multicolour photometry, the Fourier parameters and the spectroscopic 
properties of the four Cepheid candidates (see Table\,\ref{tab1}), before 
moving to the previously known Cepheid (V1154\,Cyg) in Sect.~\ref{v1154}.

\begin{figure*}
$\begin{array}{cc}
\includegraphics[angle=0,scale=0.28]{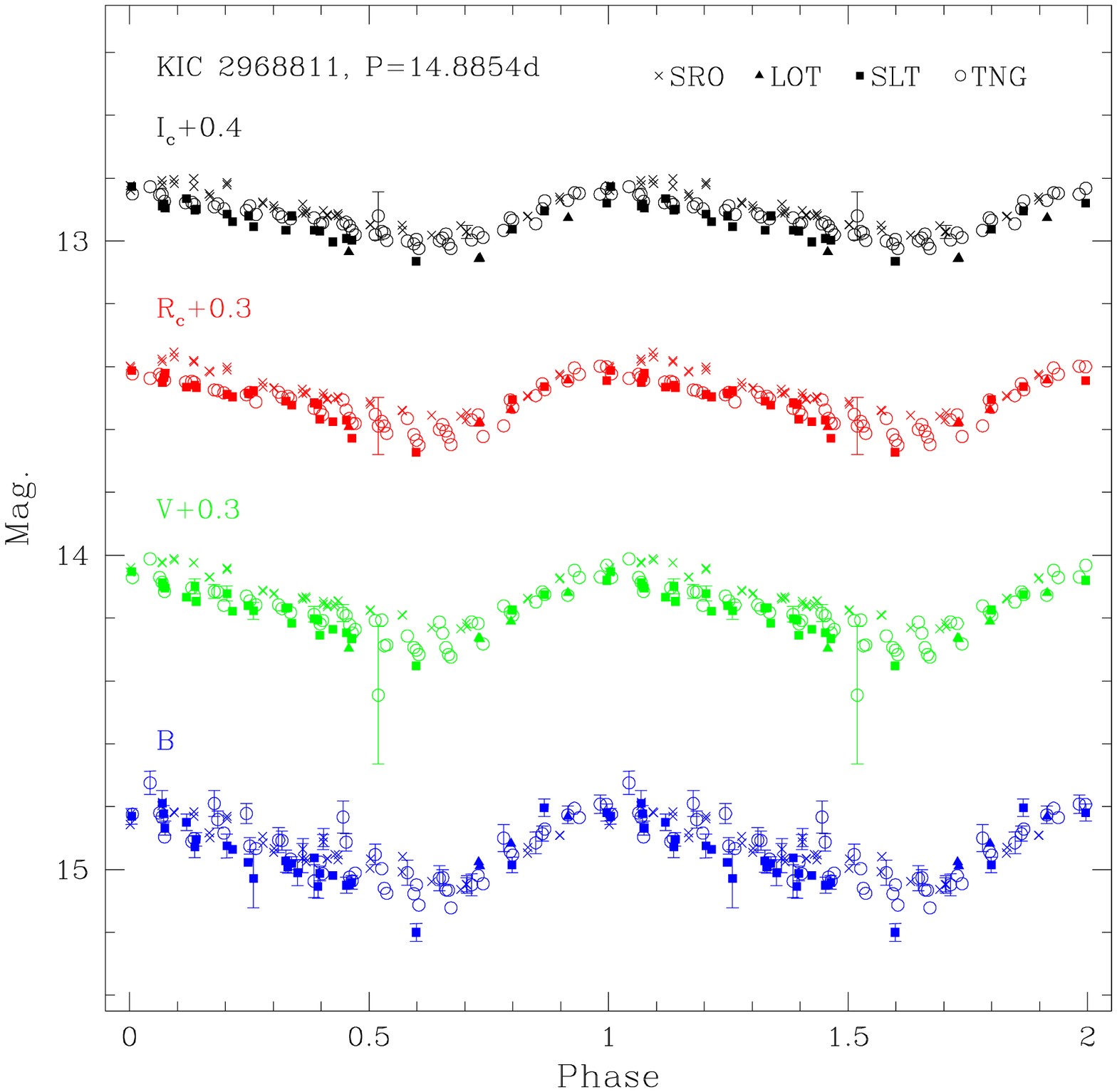}
\includegraphics[angle=0,scale=0.28]{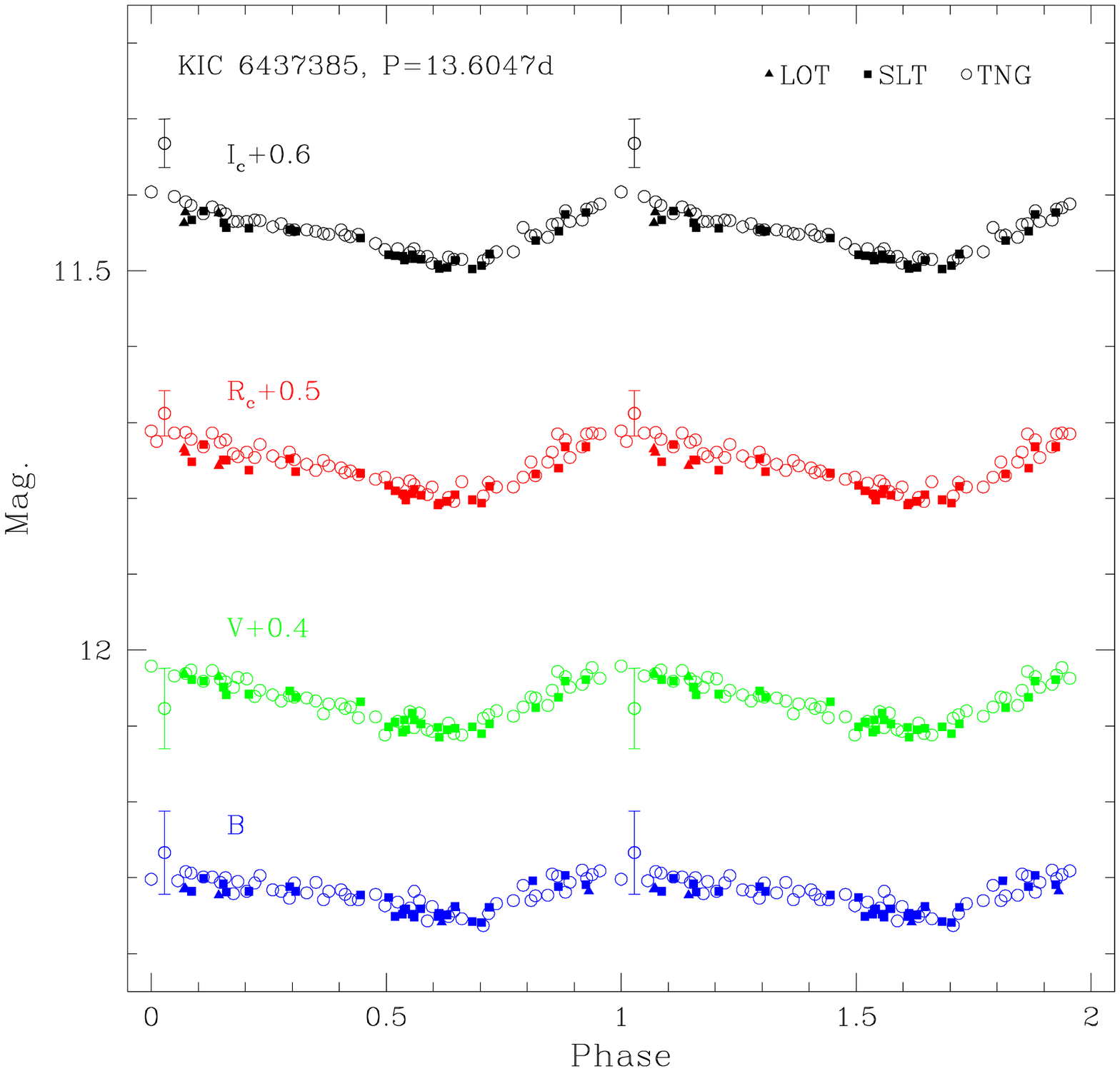} 
\includegraphics[angle=0,scale=0.28]{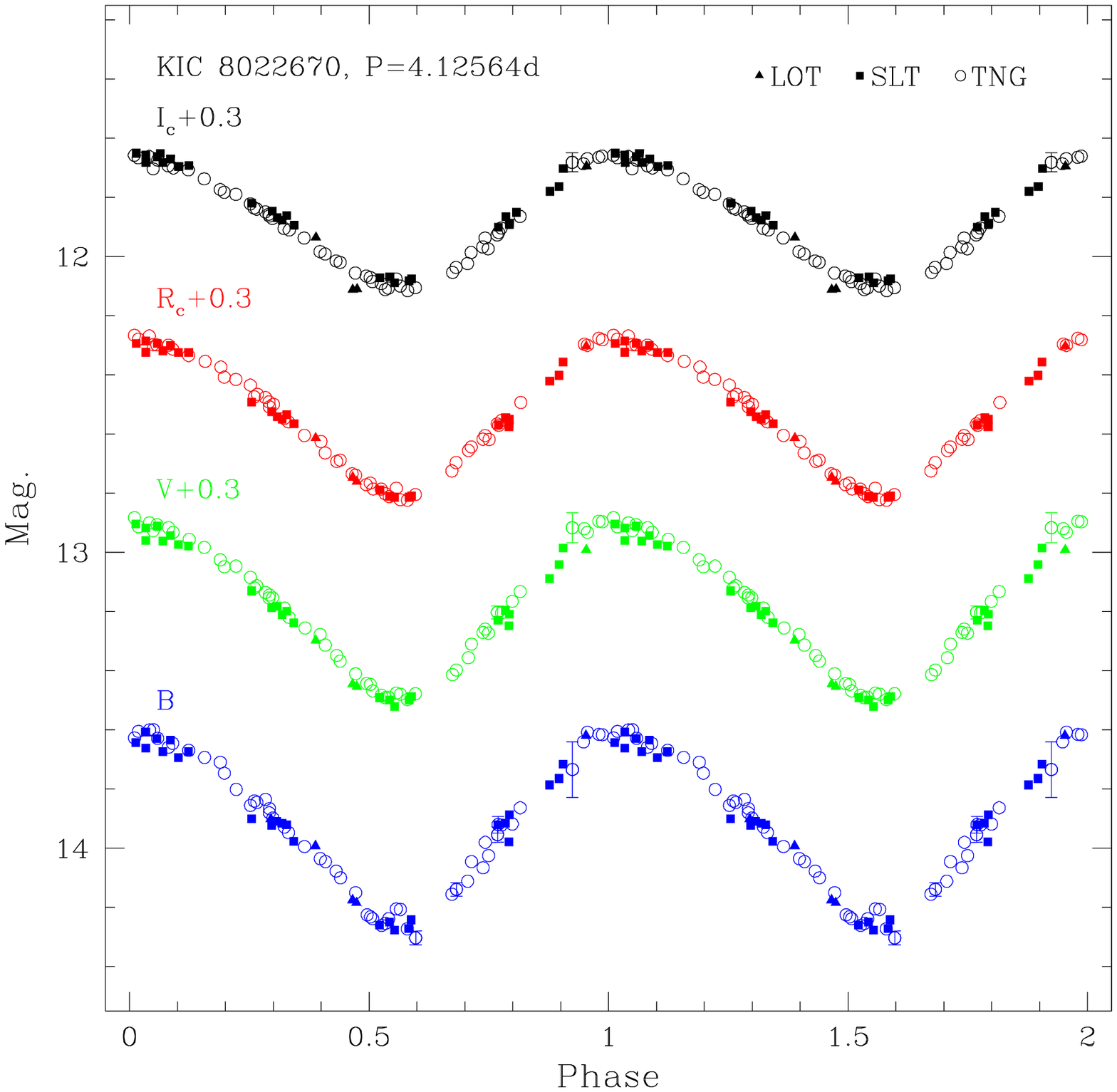} 
\end{array}$
\caption{Phased $BVR_cI_c$ Johnson-Cousins phase-folded light curves of three Cepheid candidates 
taken during the 2010 observing season: KIC\,2968811 (left panel),  KIC\,6437385 
(middle panel) and  KIC\,8022670 (right panel). For clarity, error bars were 
plotted only for data points with errors larger than 0.02\,mag. 
\label{cepcand}}
\end{figure*}

{\bf KIC\,2968811 = ASAS 190148+3807.0}  ($P = 14.8854$\,d). In principle the 
light curve shape matches all the Fourier parameter progressions within the 
uncertainty, although in the case of $\phi_{21}$ and $R_{31}$ it is a border 
line case (Fig.\,\ref{fps}). The amplitude is 0.19\,mag in the {\it Kepler} 
passband. `Shoulders' or bumps appear on the descending branch 
(Fig.\,\ref{kepphot}, upper left panel), but they are not present at the 
beginning of the observations. This effect is frequently seen in spotted stars. 
The same effect may cause the larger scatter of the folded ground-based 
multicolour light curves in the first panel of Fig.\,\ref{cepcand}. We note 
that in the case of a Cepheid in this period range the bump should be present 
on the ascending branch. By carefully examining the short brightening seen in 
the {\it Kepler} light curve we exclude instrumental or other external origin 
based on known artifacts discussed in the Kepler Data Release 
Notes\footnote{http://archive.stsci.edu/kepler/data\_release.html}. Taking into 
account the strong flare events (Fig.\,\ref{kepphot}), we conclude that this 
object is definitely not a Cepheid. 

{\bf KIC\,6437385 = ASAS 192000+4149.1} ($P = 13.6047$\,d). This star shows a
similar light curve to KIC\,2968811 with an amplitude of $0.09$\,mag
(Fig.\,\ref{kepphot} upper right panel). Within the uncertainty the Fourier 
parameters $R_{21}$ matches the progressions, but $R_{31}$ does not. The 
phases are close to the phases of the bulk of the Galactic Cepheids 
(Fig.\,\ref{fps}). The $B$ amplitude is smaller than the amplitude measured 
in $V$ as shown in Fig.\,\ref{cepcand}, this is inconsistent with the Cepheid 
classification. The light variation shows the same shoulders on the descending
branch as in the case of KIC\,2968811, while its amplitude and 
shape are changing throughout the more than 300\,d long {\it Kepler} observations.
In addition, several flares can be seen that are intrinsic to the star. One is 
shown in Fig.\,\ref{kepphot}. Thus,  KIC\,6437385 is most probably not a Cepheid.

{\bf KIC\,12406908 = ROTSE1 J192344.95+511611.8} ($P =  13.3503$\,d). 
Interestingly, this variable star exhibits very similar light curve
characteristics to the previously discussed objects (Fig.\,\ref{kepphot}, 
bottom right panel). It has an amplitude of 0.3\,mag in the {\it Kepler} 
passband. We have too few data points from ground-based observations to put 
strong constraints on the Fourier parameters, but despite the large 
uncertainties this object seems to be slightly off the main location expected 
for Cepheids (Fig.\,\ref{fps}). The ratio of $I_c$ and $V$ amplitudes is $0.74$, 
which is close to, but slightly larger, than the standard $0.6$. The SC light 
curves show small short-duration flares at BJD\,2\,455\,106.1 and 2\,455\,110.6, 
and a large one at  BJD\,2\,455\,118.2 which showed a complex, multi-peaked 
maximum lasting for 7\,h. We note that a few outliers were removed from the 
{\it Kepler} light curve, which however does not influence our conclusions. 
Based on the observed features we can safely exclude this star from our 
Cepheid sample.

{\bf KIC\,8022670 = V2279\,Cyg} ($P = 4.12564$\,d). Based on the space and 
ground-based follow-up observations this star is a strong Cepheid candidate. 
We therefore looked into the literature for a more rigorous assessment
of whether Cepheid pulsation is the cause of the observed variability.
Variability of this star was revealed by \citet{dah00} during a photographic 
search for variable stars, and independently by \citet{akerlof00} as a result 
of the ROTSE project. While \citet{akerlof00} classified the star as a Cepheid 
with a period of 4.12298\,d, \citet{dah00} interpreted it as an RS~CVn type
variable due to its proximity to a known X-ray source detected by the 
{\it ROSAT} mission \citep{vogetal99}. This object, ROTSE J191853.61+434930.0 
= LD\,349, was finally designated as V2279\,Cyg among the variable stars 
\citep{kazetal03}.

We first looked at ROTSE-I photometric data retrieved from the NSVS 
data base \citep{wozetal04}, but conclude that this data set does not 
allow us to unambiguously classify the star. A dedicated photometric 
project to select Type\,II Cepheids among the ROTSE-I targets was 
performed by \citet{schetal07}. Based on their two-colour ($V$, $R$) 
photometry they concluded that V2279\,Cyg is a probable Cepheid with 
a period of 4.117\,d. More recently, the photometric survey by 
\citet{asas09} resulted in useful data for following the period changes 
of this star. Additional photometric data are also available from the 
SuperWASP public archive\footnote{http://www.wasp.le.ac.uk/public/index.php}
and the Scientific Archive of the Optical Monitoring Camera (OMC)
on board INTEGRAL\footnote{http://sdc.laeff.inta.es/omc/index.jsp}.
Both these archives contain data obtained in a single photometric band, 
and we used these data to investigate the period behaviour of 
V2279\,Cyg.

The {\it Kepler} light curve of V2279\,Cyg seems stable, without any 
noticeable light curve changes (part of the Q2 light curve is plotted 
in Fig.\,\ref{kepphot}). We only see long term variations similar in 
amplitude to what we noted for V1154~Cyg (Fig.\,\ref{v1154lc}), which 
in this case might also just be instrumental effects. What makes 
V2279\,Cyg particularly suspicious is the presence of many flares, one 
of them is clearly seen at BJD\,2\,455\,030.0 in Fig.\,\ref{kepphot}. 
The (almost) strictly periodic variations and the value of the period 
is consistent with a rotational modulation.

Multicolour photometry can be decisive in solving this classification
problem because Cepheids have characteristic amplitude ratios. Our new 
photometric data suggest that V2279\,Cyg is not a Cepheid. The amplitude 
of its brightness variation in $V$ is only slightly smaller than the 
amplitude in the $B$ band: the ratio is about 0.9 instead of the usual 
value of $0.65-0.70$ \citep{klsz09}. Although a bright blue companion 
star is able to suppress the observable amplitude of the light variations 
in the $B$ band, the observed amplitude ratio is incompatible with the 
Cepheid nature even if the star had a very hot companion. $R_{21}$, 
$R_{31}$ and $\phi_{31}$ place it among the first overtone objects, but 
$\phi_{21}$ is very different from both fundamental and first overtone 
progressions (Fig.\,\ref{fps}). 

\begin{figure}
\includegraphics[width=83mm, angle=0]{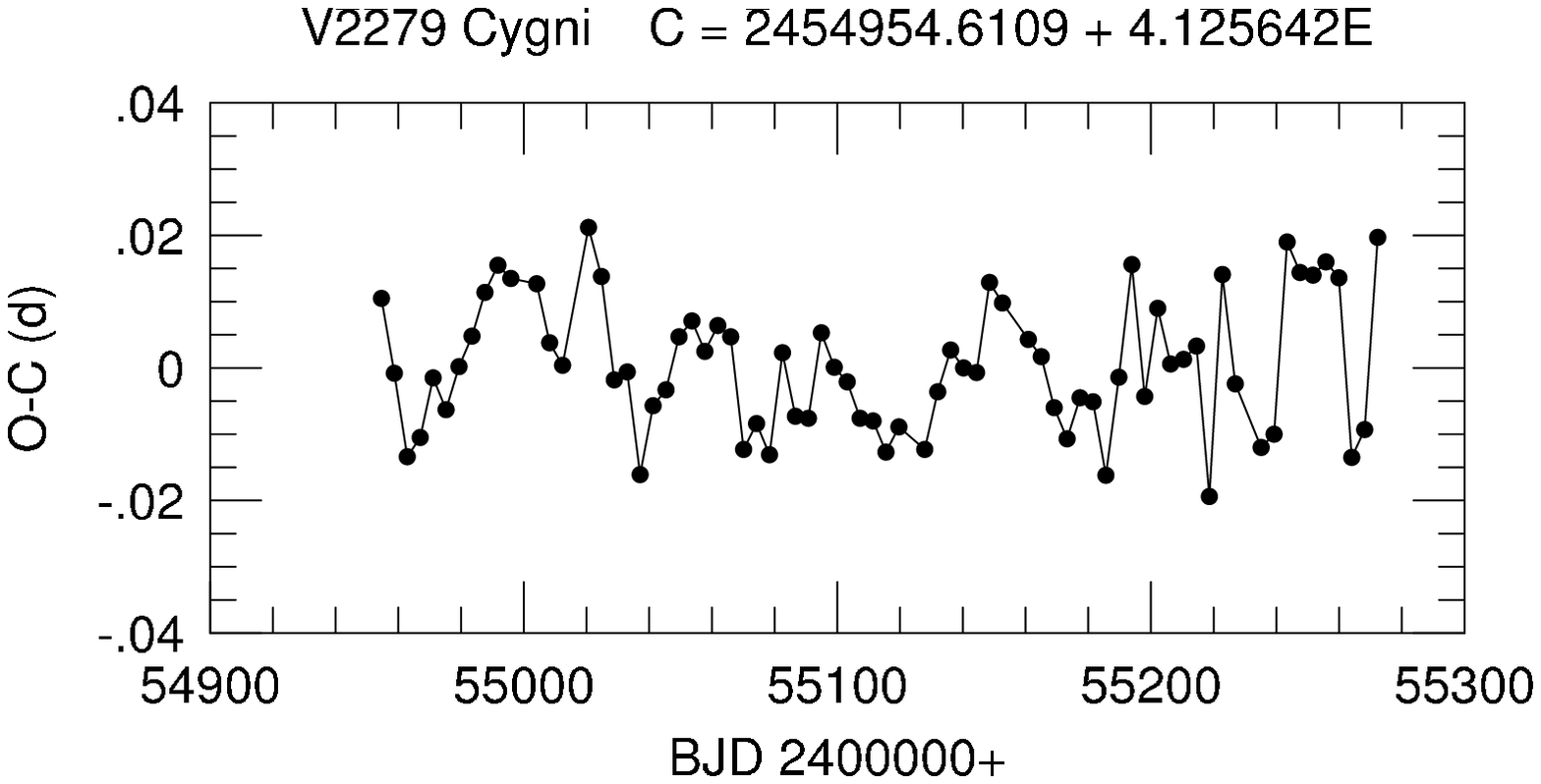}
\includegraphics[width=83mm, angle=0]{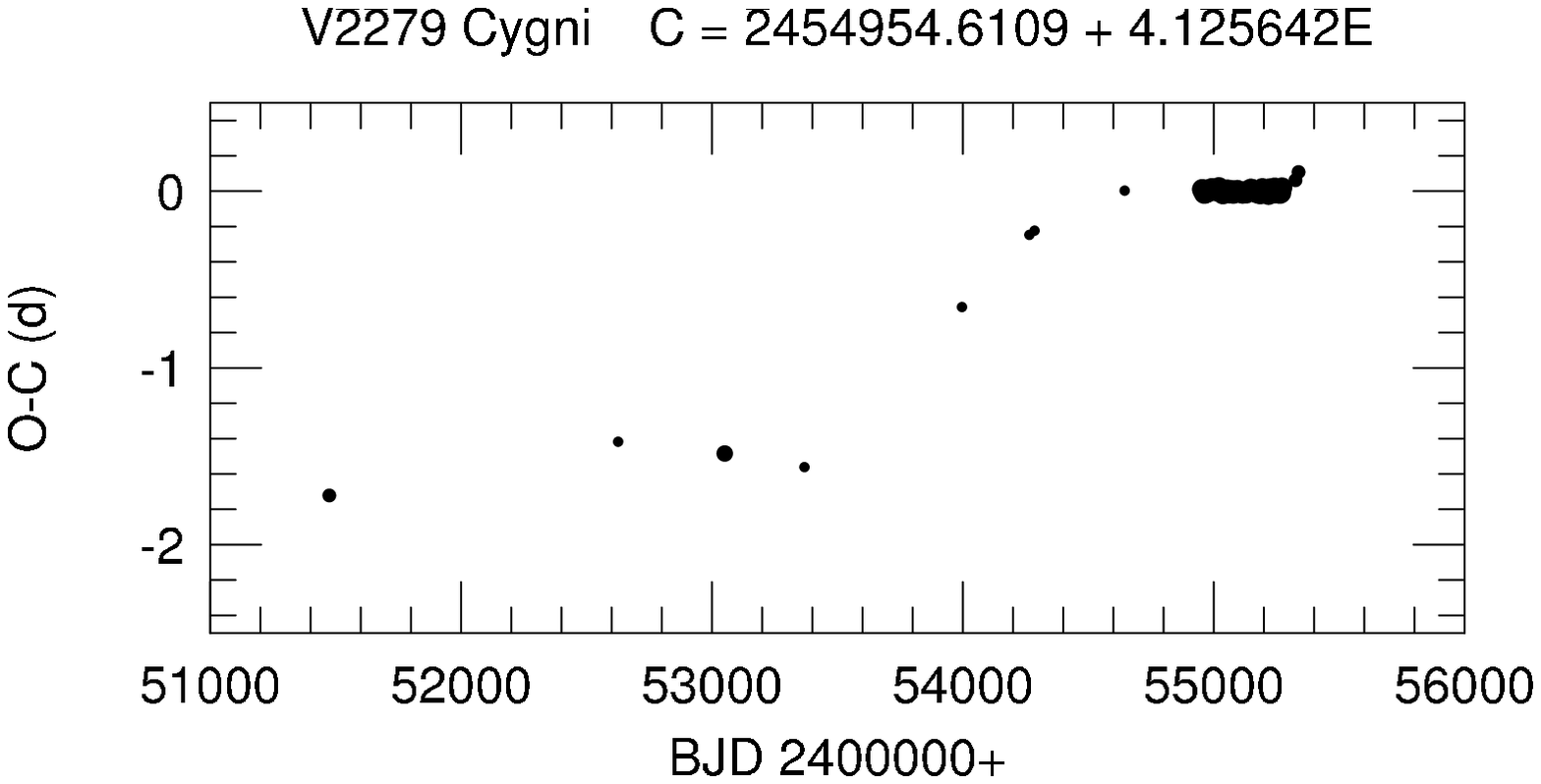}
\caption{Top panel: $O-C$ diagram of V2279\,Cyg for the {\it Kepler} 
data. Bottom panel: $O-C$ diagram of V2279\,Cyg involving all photometric 
data. The point size corresponds to the weight assigned to the maxima.}
\label{Figv2279ockepler}
\end{figure}

\begin{figure*}
\includegraphics[width=170mm, angle=0]{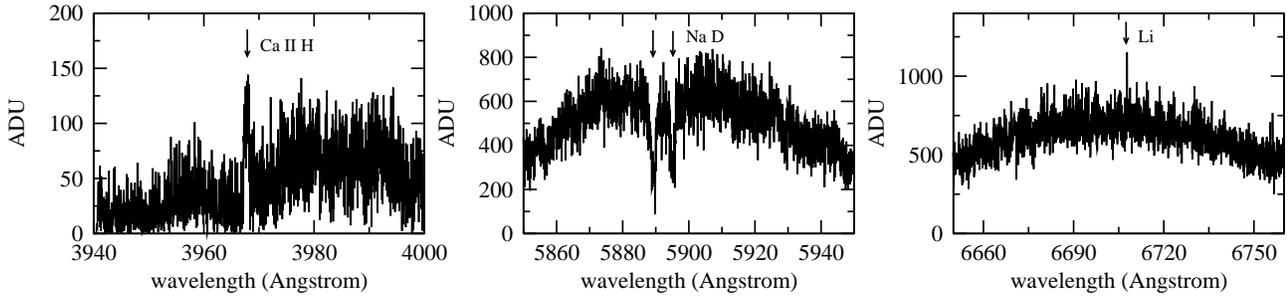}
\caption{The {\sc tres} spectrum of V2279\,Cyg. From left to right:
zoom in to Ca II H (3968.5 \AA), Na D (5890 and 5896 \AA), Li (6708 \AA).}
\label{v2279spec}
\end{figure*}

In a next step we analyze the cycle-to-cycle
variations in the periodicity using the so-called $O-C$ diagram.
The $O-C$ diagram constructed
for the moments of brightness maxima is plotted in the top panel 
of Fig.\,\ref{Figv2279ockepler}. The zero epoch was arbitrarily chosen
at the first maximum in the {\it Kepler} data set.
A least squares fit to the $O-C$ residuals resulted in the
best fitting period of 4.125642\,d, which is somewhat longer than any
formerly published value for V2279\,Cyg. We then included all ground-based
observations we could find to construct a new $O-C$ diagram using the
ephemeris

\noindent $C = {\rm BJD}\,(2\,454\,954.6109 \pm 0.0023) + (4.125642 \pm 0.000051)
E$.

\noindent Table\,\ref{TabOCv2279} contains the moments of maxima, the 
corresponding epochs and the $O-C$ residuals of the available observations.
We assigned different weights to data from different sources. These weights 
(between 1 and 5) correspond to the `goodness' of the seasonal light curve, 
larger number means better coverage and smaller scatter (for {\it Kepler} 
data W=5). The period of V2279\,Cyg show large fluctuations during the last 
decade (see the lower panel of Fig.\,\ref{Figv2279ockepler}). Our additional 
ground-based photometric data indicate a very recent change in the period 
that we will investigate using future quarters of {\it Kepler} data. We note 
that the star has a high contamination index (Table\,\ref{tab1}), 
indicating that the $\sim$18 percent of the measured flux comes form other sources. 


\begin{table}
\caption{Sample table for $O-C$ residuals for V2279 Cyg. The entire table is
available online.}
\label{TabOCv2279}
\begin{tabular}{crrcc}
\hline
\hline \noalign{\smallskip}
JD$_\odot$ $-$ & E &   $O-C$ &  W &  Reference \\
2\,400\,000  &    &[d] \ \ \ & & \\
 \noalign{\smallskip}
\hline \noalign{\smallskip}
51474.9729 & $-$843 & $-$1.7218 & 2 & NSVS \\
52626.3311 & $-$564 & $-$1.4177 & 1 & Integral OMC \\
53051.2057 & $-$461 & $-$1.4842 & 3 & Schmidt~et~al.\\
53368.6341 & $-$384 & $-$1.7304 & 1 & Schmidt~et~al.\\
54265.3816 & $-$191 & $-$1.2317 & 1 & SWASP \\
54645.2207 & $-$75  & 0.0030    & 1 & SWASP \\
54954.6214 & 0      & 0.0105    & 5 & Kepler \\
...& ...& ... & ... & ...\\
\hline
\end{tabular}
\end{table}

The spectrum of V2279\,Cyg offers the clearest evidence that this star has been 
misclassified as a Cepheid. We have plotted three segments of the spectrum taken 
with the {\sc tres} spectrograph in Fig.\,\ref{v2279spec} containing these lines: 
Ca\,\textsc{ii}\,H (3968.5\,\AA), Na\,D (5890 and 5896\,\AA) and Li (6708\,\AA). 
While the Na\,D lines are normal, the characteristics of the two other lines do 
not support a Cepheid nature of V2279\,Cyg. The Li (6708\,\AA) is never seen in 
emission in Cepheids, while the Ca emission implies chromospheric activity, which 
is also not a Cepheid characteristic. We fitted theoretical template spectra from 
the extensive spectral library of \citet{mun05} to the spectrum and determined the 
following atmospheric parameters: $T_{\rm eff}=4900\pm200$\,K, $\log{g}=3.7\pm0.4$, 
$[M/H]=-1.2\pm0.4$ and $v \sin i = 40$\,km\,s$^{-1}$. The resulting parameters are 
also incompatible with a Cepheid variable, but suggest a cool main-sequence star 
with moderate rotation.

Summarizing the previous subsections, we conclude that all the candidates turned 
out not to be Cepheids except the already known Cepheid V1154\,Cyg, which we will 
describe in detail in the following.

\section{V1154~C\lowercase{yg} the only {\it Kepler} Cepheid} \label{v1154}

In this section we analyze both {\it Kepler} 
data and ground-based follow-up observations of what is apparently the only
Cepheid being observed in {\it Kepler's} FOV.

\subsection{Observational data prior {\it Kepler}}\label{v1154-history}

Brightness variation of V1154\,Cyg was discovered by \citet{stretal63}. Cepheid 
type variation and a periodicity somewhat shorter than 5\,d were obvious from the 
photographic magnitudes leading to the discovery. The first reliable light curve 
based on photoelectric $UBV$ observations was published by \citet{wac76}. Further 
multicolour photoelectric and CCD photometric data were published by \citet{sza77},
\citet{areetal98}, \citet{igvo00}, \citet{ber08} and \citet{asas09}. This latter 
paper contains the data of a dedicated photometric survey of the whole Kepler field.
Space photometric data of V1154\,Cyg are also available from the Hipparcos satellite 
\citet{esa97} and the Scientific Archive of the Optical Monitoring Camera (OMC) on 
board INTEGRAL. None of these previous data can compete with {\it Kepler} in 
photometric quality.

\begin{figure}
\includegraphics[height=82mm, angle=0]{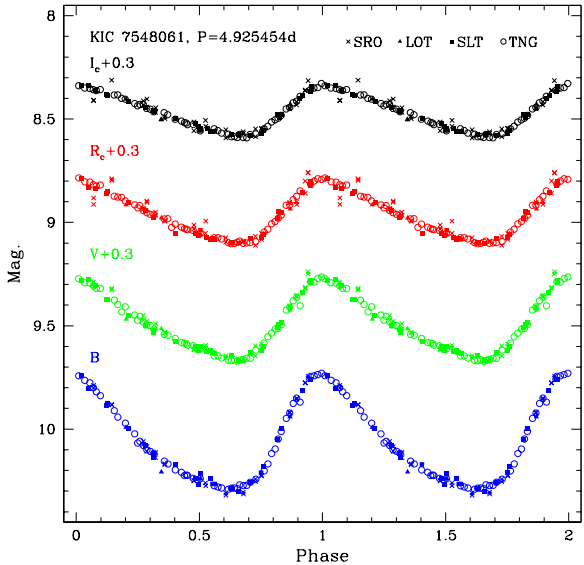}
\caption{$BVR_cI_c$ Johnson-Cousins phase-folded light curves of V1154\,Cyg taken 
during the 2010 observing season. Typical errors of the individual photometric points 
are a few mmag.}
\label{BVRI_V1154}
\end{figure}

In addition, a large number of radial velocity data have been 
collected on V1154\,Cyg by the Moscow CORAVEL team \citep{goretal98}. These data 
obtained between $1990-1996$ show a slight change in the mean 
velocity averaged over the pulsation cycle, thus \citet{goretal96} suspected 
spectroscopic binarity of this Cepheid. However, radial velocity data obtained by 
\citet{imb99} much earlier than the Moscow data and covering a reasonably long 
time interval do not indicate binarity.

\citet{mfl08} determined basic parameters for our target: 
${\rm [Fe/H]} = 0.06 \pm 0.07$, spectral type: G2Ib, $T_{\rm eff}=5370\pm118$\,K, 
$\log{g}=1.49\pm0.34$, and $v \sin i = 12.3\pm 1.6$\,km\,s$^{-1}$. These parameters 
are all consistent with a Cepheid, and place V1154\,Cyg inside the theoretical 
instability strip presented in Fig.\,\ref{KIC}. The chemical composition of 
V1154\,Cyg was determined independently by \citet{lucketal06} in a major project 
of Cepheid spectroscopy. They published a value of ${\rm [Fe/H]} = - 0.10$. 

\begin{figure}
\includegraphics[height=82mm, angle=0]{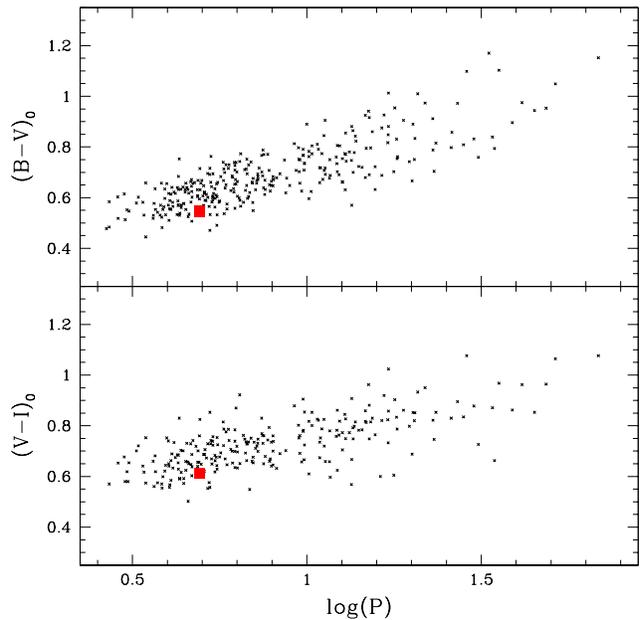}
\caption{Comparison of the extinction-corrected colours of V1154\,Cyg (filled
squares) to the Galactic Cepheids adopted from \citet{tsr03} (crosses).}
\label{PC}
\end{figure}

\subsection{Multicolour photometry of V1154\,Cyg}\label{V1154-mc}

Fig.\,\ref{BVRI_V1154} shows a multicolour light curve for V1154\,Cyg. It contains 
$114-120$ points in each filter obtained with four different telescopes as 
described in Section\,~\ref{mulcol}. The amplitudes and amplitude ratios are 
consistent with V1154\,Cyg being a Cepheid, e.g. the ratio of the $I$ and $V$ 
amplitude is 0.6, which is exactly what is expected. Fig.\,~\ref{fps} shows that 
the average Fourier parameters of V1154\,Cyg fit well all the progressions, although 
in each case they are slightly lower than the main progression. 

As a check, the colours of V1154\,Cyg were compared to other Galactic Cepheids in 
Fig.\,\ref{PC}. The observed colours of $(B-V)=0.865$ and $(V-I)=1.021$ were 
derived from the $BVR_cI_c$ light curves. To correct for extinction, we adopted 
the colour excess from 
\citet{feb95}\footnote{{http://www.astro.utoronto.ca/DDO/research/cepheids/ \\/table\_colourexcess.html}} 
(after removing the systematic trend in Fernie system using the 
prescription given in \citealt{tsr03}). The extinction-free colours are 
$(B-V)_0=0.546$ and $(V-I)_0=0.612$. As shown in Fig.\,\ref{PC}, 
these colours fit well within the period-colour relations defined by the
Galactic Cepheids.

\subsection{Analysis of the {\it Kepler} light curves} \label{KLC}

With almost a year of continuous data, it is in principle possible to study the 
stability of the light curve of a classical Cepheid over several dozen pulsation 
cycles. However, because instrumental effects are still present in the data, 
it is too early to perform such an analysis at least until pixel-level data 
become available, which would allow the data reduction to be optimized for 
this particular type of star.  

\begin{figure}
\includegraphics[width=83mm, angle=0]{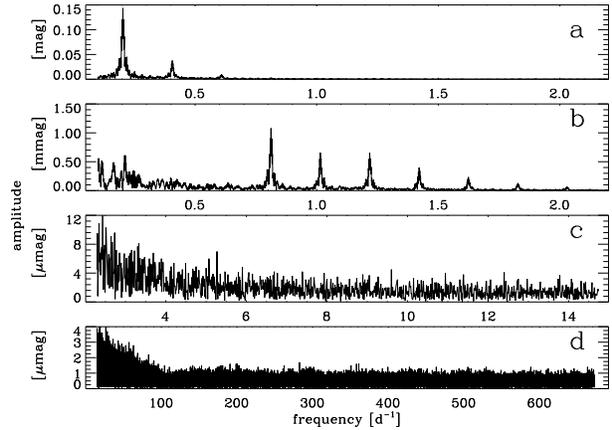}
\caption{Frequency spectrum of V1154\,Cyg based on Q1$-$Q4 LC data.
a) Main pulsational frequency and the first two harmonics. b) Prewhitened 
by the three frequency peaks, higher order harmonics emerge up to $10f_0$. 
The remaining part of the higher-frequency range of the spectrum was divided 
into two for clarity: panel c) shows the frequency range [2.4-15] $d^{-1}$, 
and panel d) comprises the range [15-700] $d^{-1}$ based on Q5 SC data. Note 
the change of the scale on the vertical axis.}
\label{w4}
\end{figure}

The frequency content of the light curve of V1154\,Cyg was investigated with 
standard Fourier transform methods by applying well-tested software packages:
{\sc SigSpec} \citep{reg07}, {\sc Period04} \citep{lb05} and {\sc MuFrAn} 
\citep{kol90}. The distorted parts of the LC light curves have been omitted, 
because their presence causes spurious frequency peaks around the main frequencies. 
The affected parts are the entire Q0 and ${\rm BJD}=[2455054.0-2454094.24]$ in Q2. 
The frequency spectrum shows the main pulsation frequency at $f_0=0.203$\,d$^{-1}$ 
and many harmonics. Two harmonics ($2f_0$ and $3f_0$) are clearly visible in the 
upper panel of Fig.\,~\ref{w4}. Prewhitening with these peaks reveals further 
harmonics up to the $10^{\rm th}$ order with very low amplitudes. This is the first 
time that such high-order harmonics have been detected, underlining the accuracy of 
the {\it Kepler} observations. The effect of instrumental artefacts (trends, 
amplitude variation) is clearly seen in the remaining power around $f_0$. Apart 
from that the frequency spectrum is completely free of additional power at the 
significance level up to the Nyquist frequency as shown in the lower two panels 
of Fig.\,\ref{w4}. 

Before finishing this paper, Q5 SC data became available for V1154\,Cyg.
To investigate the high-frequency range we used this 94.7\,d long data set
and the 33.5\,d long SC data taken in Q1. The two data sets have a very similar
frequency content, and we chose to plot the Q5 SC frequency spectrum in 
panel d of Fig.\,\ref{w4}, because the longer timebase ensures better frequency
resolution and higher SNR. 

The top of the grass of the remaining spectrum is 5\,$\mu$mag, 
while the average is 2\,$\mu$mag in the spectrum up to 50\,d$^{-1}$. Above that 
the top of the grass of the remaining spectrum decreases to 1.5\,$\mu$mag, and from 
100\,d$^{-1}$ it remains constant. 
The average of the remaining peaks 
is below 1\,$\mu$mag in this high-frequency range. The residual spectrum shows no 
signal of any shorter period nonradial pulsation modes or solar-like oscillations.

The frequencies, amplitudes and phases of the detected and identified peaks 
based on Q1$-$Q4 LC data are listed in Table\,\ref{tabfreq}. The zero epoch was 
chosen close to moment of the first data point, i.e. ${\rm BJD}=2454954.0$. The 
errors have been estimated from Period04. Searching for only one frequency of the 
highest amplitude at a time and prewhitening for it and then repeating the 
procedure gave practically the same results as searching for all the harmonics 
simultaneously.

\begin{table}
\caption{Frequencies, amplitudes and phases of the identified 
   frequency peaks in the frequency spectrum of V1154\,Cyg based on Q1$-$Q4 
   LC data. 
   The formal uncertainty (1$\sigma$) of the amplitudes is uniformly 
   $0.000024$.}
\label{tabfreq}
\begin{tabular}{@{}rccccc}
\hline\hline
     ID &  freq.    & $\sigma_f$ & ampl. &  $\varphi$ phase & $\sigma_{\varphi}$ \\
        &  d$^{-1}$ & d$^{-1}$   & mag   &  rad             & rad \\
  \hline
 $f_0$  & 0.2030244  & 0.0000002 & 0.144984 & 0.794350 & 0.000027\\
 $2f_0$ & 0.4060514  & 0.0000010 & 0.039291 & 4.403636 & 0.000097\\
 $3f_0$ & 0.6090704  & 0.0000041 & 0.009880 & 1.926416 & 0.000385\\
 $4f_0$ & 0.8120798  & 0.0000375 & 0.001090 & 5.812189 & 0.003496\\
 $5f_0$ & 1.0151798  & 0.0000668 & 0.000611 & 5.508786 & 0.006232\\
 $6f_0$ & 1.2181468  & 0.0000683 & 0.000598 & 3.132366 & 0.006361\\
 $7f_0$ & 1.4211946  & 0.0001077 & 0.000379 & 0.442772 & 0.010040\\
 $8f_0$ & 1.6241924  & 0.0001820 & 0.000224 & 4.130455 & 0.016973\\
 $9f_0$ & 1.8272617  & 0.0003652 & 0.000112 & 1.421814 & 0.034039\\
$10f_0$ & 2.0304263  & 0.0006842 & 0.000060 & 4.893997 & 0.063769\\
$11f_0$ & 2.2334364  & 0.0001591 & 0.000026 & 2.184016 & 0.148302\\
\hline
\end{tabular}
\end{table}

\subsection{Behaviour of the pulsation period}\label{v1154-period}

This is the first occasion that cycle-to-cycle changes in the
pulsation period of a Cepheid can be followed. The $O-C$ analysis
of {\it Kepler} data of V1154\,Cyg is published in a separate paper
(Derekas et al. 2011, in prep.). Here we study the long-term behaviour of the
pulsation period.

The computed times of maxima were calculated from the period fitted 
to the {\it Kepler} data. One $O-C$ point was derived for the 
mid-epoch of annual sections for each available photometric 
time series taken from the literature. Besides the {\it Kepler} maxima 
and available CCD and photoelectric observation we publish for the first 
time V1154\,Cyg data from digitized Harvard plates and eye estimations
from Sternberg Astronomical Institute (SAI) photographic plates. The 
passband of these observations is close to Johnson $B$. We also used 
visual and CCD observations from the AAVSO International Database. Where 
both Johnson $B$ and $V$ data were available at the same epoch, we 
retained only the $V$ maxima.

These data points are listed in the first column of 
Table\,\ref{TabOC}. Column\,2 lists the epoch number ($E=0$ was 
arbitrarily taken at the first maximum of the {\it Kepler} data). 
Weights were assigned to individual data sets 
similarly to the case of V2279\,Cyg. The weights are listed in 
Column\,4. The final ephemeris was derived by a weighted linear least 
squares fit to the preliminary residuals computed from a formerly published 
(usually slightly incorrect) period value which is an inherent step 
in the $O-C$ method. No weight was assigned to the photographic 
normal maxima, these $O-C$ residuals were omitted from fitting procedure.
The $O-C$ residuals in Column\,3 have been calculated with the 
final ephemeris

\noindent $C = {\rm BJD}\,(2\,454\,955.7260 \pm 0.0008) + (4.925454 \pm 
0.000001)\,{\rm E}$. 

\begin{figure}
\includegraphics[width=83mm, angle=0]{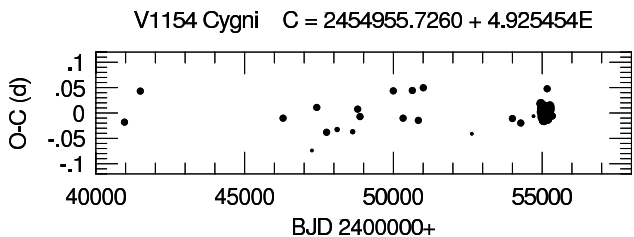}
\caption{$O-C$ diagram of V1154\,Cyg. This plot indicates that the
pulsation period has been constant since 1970. The $O-C$ residuals
obtained from the {\it Kepler} data (rightmost clump of points)
are analyzed in a separate paper (Derekas et al. 2011, in prep.).}
\label{FigOC}
\end{figure}

\noindent This period is considered to be more accurate than the 
one obtained by fitting the frequency and its harmonics. 
The source of data is given in the last column of Table\,\ref{TabOC}.
The $O-C$ diagram is shown in Fig.\,\ref{FigOC}. This plot indicates that 
the pulsation period has been constant during the last 40\,y. In principle, the 
moments of maximum light differ for different passbands. From simultaneous 
observations we find a difference of $-0.020$\,d between the $V$ and $Kp$ maxima 
in the sense of $V - Kp$, which is neglected in Fig.\,\ref{FigOC}. We note that 
this does not change our conclusions in any way. 

\subsection{Radial velocities of V1154\,Cyg}\label{V1154-bin}

\begin{figure}
\includegraphics[width=83mm, angle=0]{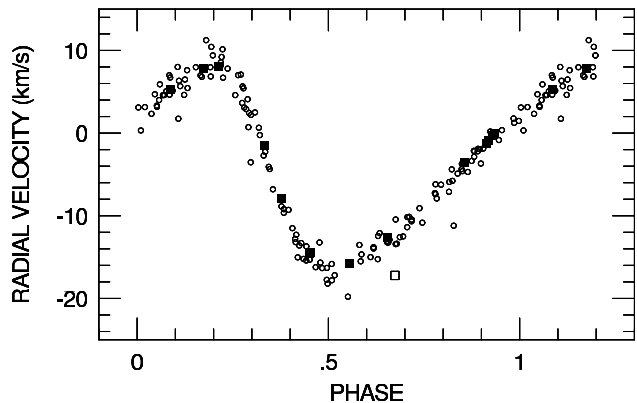}
\caption{Phased radial velocity data of V1154\,Cyg.
Open circles denote data obtained by \citet{goretal98},
and \citet{imb99}, the new Tautenburg data appear as filled squares, 
while the INAF-OACt spectrum is denoted by an open square.}
\label{Fig-RVcomp}
\end{figure}

The suspected spectroscopic binarity of V1154\,Cyg can be 
investigated with the help of the new radial velocity data, most of them 
obtained with the Tautenburg 2.0~m telescope (see Table\,\ref{tabspec}).
The comparison with the data taken from the literature \citep{goretal98}
and \citet{imb99} does not give a new evidence of spectroscopic binarity
because the $\gamma$-velocity (mean radial velocity averaged over a
complete pulsational cycle) derived from the new data practically
coincides with that obtained from the previous observations -- see
Fig.\,\ref{Fig-RVcomp}.
Quantitatively, the difference is $-0.19 \pm 0.30$\,km\,s$^{-1}$ and
$-0.15 \pm0.22$\,km\,s$^{-1}$ between our new data and data of 
\citet{goretal98} and \citet{imb99}, respectively.

We add that a period derived from all the available radial velocity data 
is in very good agreement with (less than 2-$\sigma$ from) the photometric 
period derived in the previous subsection.


\begin{table}
\caption{Sample table of $O-C$ residuals for V1154\,Cyg. 
The entire table is available online.}
\label{TabOC}
\begin{tabular}{c@{\hskip2mm}c@{\hskip2mm}r@{\hskip2mm}c@{\hskip2mm}c@{\hskip2mm}c@{\hskip2mm}}
\hline
\hline \noalign{\smallskip}
JD$_\odot$ $-$ & E &   $O-C$ &  W & filter & Reference \\
2\,400\,000  &    &    [d] \ \ \ & & & (Instrument)\\
 \noalign{\smallskip}
\hline \noalign{\smallskip}

14862.215 &  $-$8140 & $-$0.3154 & 0 &   PG & this work (Harvard) \\
15675.172 &  $-$7975 & $-$0.0584 & 0 &   PG & this work (Harvard) \\
16492.860 &  $-$7809 &    0.0043 & 0 &   PG & this work (Harvard) \\
 ... & ... & ...  & ... & ... & ...\\
40962.4929 & $-$2841 & $-$0.0188  & 3 & V &\citet{wac76} \\
41494.5033 & $-$2733 &    0.0426  & 3 & V &\citet{sza77} \\
46291.8422 & $-$1759 & $-$0.0107  & 3 & V &\citet{ber08} \\
 ... & ... & ...  & ... & ... & ...\\
54955.7450 &     0 &  0.0185 &  5 & Kp & this work (Kepler) \\
54960.6690 &     1 &  0.0170 &  5 & Kp & this work (Kepler) \\ 		   
54965.5787 &     2 &  0.0014 &  5 & Kp & this work (Kepler) \\
\hline
\end{tabular}
\end{table}

\begin{table}
\begin{center}
\caption{Log of the spectroscopic observations of V1154\,Cyg containing the 
date of observation, the computed radial velocity, the error of the 
radial velocity, the signal-to-noise ratio of the spectrum and the 
observatory. The exposure time was 30 minutes in each case.}
\label{tabspec}
\begin{tabular}{crccc}
\hline
 JD$_{\odot}$    &    RV\ \   &  $\sigma$ & SNR  & Obs. \\ 
2\,450\,000+  & km\,s$^{-1}$  & km\,s$^{-1}$ & &\\ \hline
4267.4519 &$-$17.22  &  0.7   &  60 & OACt \\
5338.5157 &    8.069 &  0.033 & 193 & TLS \\
5359.3851 &$-$14.597 &  0.053 &  72 & TLS \\
5376.4405 & $-$1.216 &  0.006 & 107 & TLS \\
5376.4621 & $-$0.921 &  0.006 & 110 & TLS \\ 
5378.5110 & $-$1.540 &  0.006 & 110 & TLS \\
5381.4465 & $-$0.243 &  0.007 &  91 & TLS \\
5381.4681 & $-$0.076 &  0.007 &  96 & TLS \\
5389.4482 &$-$15.779 &  0.046 &  74 & TLS \\ 
5393.5050 & $-$7.969 &  0.019 &  93 & TLS \\
5397.4266 &    7.848 &  0.040 & 214 & TLS \\
5428.3559 &$-$14.464 &  0.048 &  80 & TLS \\ 
5429.3496 &$-$12.620 &  0.048 &  75 & TLS \\
5430.3332 & $-$3.570 &  0.015 & 101 & TLS \\
5431.4743 &    5.302 &  0.019 & 141 & TLS \\
\hline
\end{tabular}
\end{center}
\end{table}

\subsection{Pulsation mode of V1154\,Cyg}\label{mode}

Cepheids pulsate in one of the first three radial modes (fundamental (F), first 
(O1) and second overtone (O2)) or simultaneously in two or three of them. Some
triple-mode Cepheids pulsate in the first three overtones at the same time. 
From a pulsational and evolutionary point of view it is important to
determine the pulsational mode of a monoperiodic Cepheid. 
Cepheids with pulsational periods similar to V1154\,Cyg may pulsate in 
the fundamental or the first overtone mode. The usual way of distinction 
is the use of Fourier parameters that show characteristic progression 
as a function of period. However, the Fourier parameters of radial 
velocity curves are indistinguishable for fundamental and first overtone 
pulsators with periods around 5\,d (see Fig.\,3 of \citealt{bsd09}).
Light curve Fourier parameters suffer from similar problems. 
Based on Fig.\,\ref{fps}, $R_{21}$ and $\phi_{31}$ are the most 
promising parameters for mode discrimination. However, a closer 
inspection reveals that there is a $2\pi$ difference between $\phi_{31}$
values of fundamental and first overtone Galactic Cepheids close to 5\,d period, 
therefore this particular phase difference is not a good discriminator in the 
case of V1154\,Cyg. $R_{21}$ is higher for F Cepheids, and lower for O1
Cepheids. However it is amplitude dependent, and if we decrease the 
amplitude of the F Cepheid it also goes to zero. V1154\,Cyg is not a 
low-amplitude Cepheid, and based on $R_{21}$ alone, it can be classified 
as a fundamental mode pulsator. Though to draw a firm conclusion, we need more 
pieces of evidence. 

\begin{figure}
\includegraphics[width=83mm, angle=0]{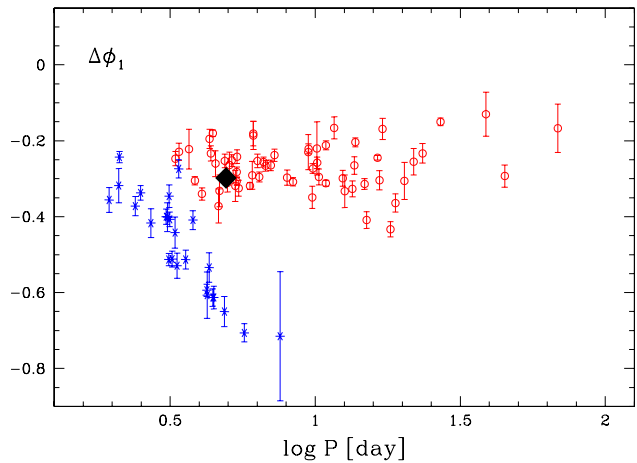}
\caption{Phase lag of the fundamental (red circles) and first overtone 
(blue asterisks) Galactic Cepheids. V1154\,Cyg is denoted by a black diamond.}
\label{phaselag}
\end{figure}

Analysis of the first order phase lag ($\Delta \Phi_1 = \phi^{V_r}_1 - \phi^{V}_1 $) 
between the Johnson $V$ light curve and radial velocity may come to the 
rescue \citep{omk00, sbb07}. The phase lag method is a reliable tool to 
establish Cepheid pulsation mode in this pulsation period regime. We used 
the simultaneous new RV data (Table\,\ref{tabspec}) and Johnson $V$ 
photometry (Fig.\,\ref{BVRI_V1154} and Table\,\ref{onl1}) and derived 
$\Delta \Phi_1 = -0.298 \pm 0.018$ which places our Cepheid firmly 
among fundamental mode Cepheids as is clearly seen in Fig.\,\ref{phaselag}.


\section[]{Conclusions}\label{disc}

We described the pre-launch selection of Cepheid candidates within the 
{\it Kepler} FOV. Of our forty candidates only five remained after inspection
of the {\it Kepler} light curves. Aided by additional ground-based multicolour 
and spectroscopic observations we excluded further four stars including
V2279\,Cyg (KIC\,8022670), which is most likely a rapidly rotating K dwarf
with flares showing prominent emission lines. Our results show that its
previous classification as a Cepheid is not correct. This leaves only
one star, V1154\,Cyg KIC\,7548061), which is a well-known Cepheid.

{\it Kepler} has provided one of the most accurate Cepheid light curves to 
date. High-order harmonics of the main pulsational mode were detected for 
the first time up to the $10^{\rm th}$ harmonic. Reliable investigation of 
cycle-to-cycle variations in the light curve is currently hampered by 
instrumental effects, but will be investigated as pixel-level data or 
finally corrected data are available for V1154\,Cyg. Period variation is 
investigated in a separate paper (Derekas et al. 2011, in prep).

New, high-precision radial velocity measurements of V1154\,Cyg do not 
confirm spectroscopic binarity hypothesized by \citet{goretal96}. 
Measuring the phase lag between simultaneous photometric and radial
velocity data of the pulsation allowed us to determine unambiguously 
that V1154\,Cyg is a fundamental mode pulsator.

An intriguing feature of classical Cepheids is the possible presence of 
nonradial pulsation modes in the case of a number of first-overtone Cepheids
in the Large Magellanic Cloud \citep{mk09}. Indeed, theoretical computations by
\citet{mmb07} predict the presence of high-order nonradial mode close to and 
beyond the blue edge of the Cepheid instability strip. 
Although V1154\,Cyg pulsates in the fundamental mode, we have searched for, but 
have not found any short period variability (nonradial or stochastically excited 
modes).

New kinds of investigations will become possible with upcoming years-long 
{\it Kepler} data when all instrumental effects are understood. 
These include analyses of the light curve variation from cycle to cycle, 
and the detection of low-mass companions through the light-time effect.

\section*{Acknowledgments}

Funding for thr {\em Kepler Mission} is provided by NASA's Science 
Mission Directorate. This project has been supported by the `Lend\"ulet' 
program of the Hungarian Aca\-demy of Sciences and the Hungarian OTKA grant K83790. 
Financial support from the ESA (PECS programme No.\,C98090) is gratefully 
acknowledged. CCN thanks the funding from National Science Council (of Taiwan) 
under the contract NSC 98-2112-M-008-013-MY3. RS is supported by the 
Austrian Science Fund (FWF project AP 2120521). We acknowledge 
assistance of the queue observers, Chi-Sheng Lin and Hsiang-Yao 
Hsiao, from Lulin Observatory. Part of this project is based on 
data from the OMC Archive at LAEFF, pre-processed by ISDC. We 
acknowledge with thanks the variable star observations from the 
AAVSO International Database contributed by observers worldwide 
and used in this research. The authors gratefully acknowledge the 
entire Kepler team, whose outstanding efforts have made these 
results possible.

\label{lastpage}


\begin{thebibliography}{99}

\bibitem[\protect\citeauthoryear{Akerlof et~al.}{2000}]{akerlof00} Akerlof C., 
Amrose S., Balsano R. et al., 2000, AJ, 119, 1901

\bibitem[\protect\citeauthoryear{Antonello \& Lee}{1981}]{fp1} Antonello E., 
Lee A.~S., 1981, ApJ, 248, 291

\bibitem[\protect\citeauthoryear{Antonello \& Morelli}{1996}]{fp7} Antonello E., 
Morelli P.~L., 1996, A\&A, 312, 541

\bibitem[\protect\citeauthoryear{Antonello \& Poretti}{1986}]{fp3} Antonello E., 
Poretti E., 1986, A\&A, 169, 149

\bibitem[\protect\citeauthoryear{Antonello, Poretti \& Reduzzi}{1990}]{fp4} Antonello E., 
Poretti E., Reduzzi L., 1990, A\&A, 236, 138

\bibitem[\protect\citeauthoryear{Arellano Ferro et~al.}{1998}]{areetal98}
Arellano Ferro A., Rojo Arellano R., Gonz\'alez-Bedolla S.,              
Rosenzweig P., 1998, ApJS, 117, 167                    

\bibitem[\protect\citeauthoryear{Baranowski et~al.}{2009}]{bsd09} Baranowski R., 
Smolec R., Dimitrov W. et~al., 2009, \linebreak MNRAS, 396, 2194

\bibitem[\protect\citeauthoryear{Batalha et~al.}{2010}]{bbk10} Batalha N.~M., 
Borucki W.~J., Koch D.~G. et~al., 2010, ApJ, 713, L109

\bibitem[\protect\citeauthoryear{Berdnikov}{2008}]{ber08}
Berdnikov L.~N., 2008, VizieR On-line Data Catalog: II/285

\bibitem[\protect\citeauthoryear{Berdnikov}{2010}]{berd10} Berdnikov L.~N., 
2010, in {Variable Stars, the Galactic Halo and Galaxy Formation}, Eds.:
{C.~Sterken, N.~Samus, \& L.~Szabados}, Moscow, Sternberg Astr. Inst.
 
\bibitem[\protect\citeauthoryear{Bertin \& Arnouts}{1996}]{BA96} Bertin, E., 
Arnouts, S., 1996, A\&AS, 117, 393

\bibitem[\protect\citeauthoryear{Blomme et~al.}{2010}]{BDDR10} Blomme J., 
Debosscher J., De Ridder J. et~al., 2010, ApJ, 683, 433

\bibitem[\protect\citeauthoryear{Borucki et~al.}{2010}]{BKB10} Borucki W.~J., 
Koch D., Basri G. et~al., 2010, Science, 327, 977

\bibitem[\protect\citeauthoryear{Bruntt et~al.}{2008}]{bes08} Bruntt H., 
Evans N.~R., Stello D. et~al., 2008, ApJ, 683, 433

\bibitem[\protect\citeauthoryear{Burki et~al.}{1986}]{BSA86} Burki G., 
{Schmidt} E.~G., {Arellano Ferro} A., {Fernie} J.~D., {Sasselov} D., {Simon}
N.~R., {Percy} J.~R., {Szabados} L., 1986, A\&A, 168, 139

\bibitem[\protect\citeauthoryear{Dahlmark}{2000}]{dah00}
Dahlmark L., 2000, IBVS, No.\,4898


\bibitem[\protect\citeauthoryear{ESA}{1997}]{esa97}
ESA 1997, Hipparcos Catalogue, ESA SP-1200         

\bibitem[\protect\citeauthoryear{Fernie et~al.}{1995}]{feb95} Fernie J.~D., 
Evans N.~R., Beattie B., Seager, S., 1995, IBVS, 4148, 1

\bibitem[\protect\citeauthoryear{Gilliland et~al.}{2010a}]{gil10a} Gilliland
R.~L., Brown T.~M., Christensen-Dalsgaard J. et~al., 2010a, PASP, 122, 131

\bibitem[\protect\citeauthoryear{Gilliland et~al.}{2010b}]{gil10b} Gilliland
R.~L., Jenkins J.~M., Borucki W.~J. et~al., 2010b, ApJ, 713, L160

\bibitem[\protect\citeauthoryear{Gorynya et~al.}{1996}]{goretal96}
Gorynya N.~A., Samus' N.~N., Rastorguev A.~S., Sachkov M.~E., 
1996, AstL, 22, 175

\bibitem[\protect\citeauthoryear{Gorynya et~al.}{1998}]{goretal98}
Gorynya N.~A., Samus' N.~N., Sachkov M.~E., Rastorguev A.~S.,
Glushkova E.~V., Antipin S.~V., 1998, AstL, 24, 815

\bibitem[\protect\citeauthoryear{Hartman et~al.}{2004}]{hat04} Hartman J.~D., 
Bakos G., Stanek K.~Z., Noyes R.~W., 2004, AJ, 128, 1761

\bibitem[\protect\citeauthoryear{Ignatova \& Vozyakova}{2000}]{igvo00}
Ignatova V.~V., Vozyakova O.~V., 2000, Astron. Astrophys. Trans., 19, 133

\bibitem[\protect\citeauthoryear{Imbert}{1999}]{imb99} Imbert M., 1999, A\&AS, 140, 79

\bibitem[\protect\citeauthoryear{Jenkins et~al.}{2010}]{jen10a} Jenkins J.~M.,
Caldwell D.~A., Chandrasekaran H. et~al., 2010a, ApJ, 713, L87 

\bibitem[\protect\citeauthoryear{Jenkins et~al.}{2010}]{jen10b} Jenkins J.~M.,
Caldwell D.~A., Chandrasekaran H. et~al., 2010b, ApJ, 713, L120 

\bibitem[\protect\citeauthoryear{Kazarovets et~al.}{2003}]{kazetal03}
Kazarovets E. V., Kireeva N.~N., Samus N.~N., Durlevich O.~V., 2003,
IBVS, No.\,5422

\bibitem[\protect\citeauthoryear{Klagyivik \& Szabados}{2009}]{klsz09}
Klagyivik P., Szabados L., 2009, A\&A, 504, 959

\bibitem[\protect\citeauthoryear{Koch et al. }{2010}]{koch10}
Koch D.~G., Borucki W.~J. Basri G. et al., 2010, ApJL, 713, 79

\bibitem[\protect\citeauthoryear{Kolenberg et~al.}{2010}]{kk10b} Kolenberg K., 
Szab\'o R., Kurtz D.~W. et~al., 2010, \linebreak MNRAS, accepted,
arXiv:1011.5908

\bibitem[\protect\citeauthoryear{Koll\'ath}{1990}]{kol90} Koll\'ath Z., 
1990, Occ. Tech. Notes, Konkoly Obs. No. 1.

\bibitem[\protect\citeauthoryear{Landolt}{2009}]{Land09} Landolt A.~U., 
2009, AJ, 137, 4186

\bibitem[\protect\citeauthoryear{Lenz \& Breger}{2005}]{lb05} Lenz P., Breger
M., 2005, CoAst, 146, 53

\bibitem[\protect\citeauthoryear{Luck et~al.}{2006}]{lucketal06}
Luck R.~E., Kovtyukh V.~V., Andrievsky S.~M., 2006, AJ, 132, 902

\bibitem[\protect\citeauthoryear{Mantegazza \& Poretti}{1992}]{fp5} Mantegazza
L., Poretti E., 1992, A\&A, 261, 137

\bibitem[\protect\citeauthoryear{Moffett \& Barnes}{1985}]{fp2} Moffett T.~J.,
Barnes T.~G.~III, 1985, ApJS, 58, 843

\bibitem[\protect\citeauthoryear{Molenda-\.Zakowicz et al.}{2008}]{mfl08} 
Molenda-\.Zakowicz J., Frasca A., Latham D.~W., 2008, AcA, 58, 419

\bibitem[\protect\citeauthoryear{Moskalik \& Ko\l{}aczkowski}{2009}]{mk09} Moskalik
P., Ko\l{}aczkowski Z., 2009, MNRAS, 394, 1649

\bibitem[\protect\citeauthoryear{Mulet-Marquis et~al.}{2007}]{mmb07}
Mulet-Marquis C., Glatzel W., Baraffe I., Winisdoerffer C., 2007, A\&A, 465, 937

\bibitem[\protect\citeauthoryear{Munari et~al.}{2005}]{mun05} Munari, U., 
Sordo R., Castelli F., Zwitter T., 2005, A\&A, 442, 1127	

\bibitem[\protect\citeauthoryear{Og\l{}oza et al.}{2000}]{omk00} Og\l{}oza W., 
Moskalik P., Kanbur S., 2000, ASP Conf. Ser. 203, 235

\bibitem[\protect\citeauthoryear{Pigulski et~al.}{2009}]{asas09}
Pigulski A., Pojma{\'n}ski G., Pilecki B., Szczygie{\l}, D.~M., 2009, AcA,
59, 33

\bibitem[\protect\citeauthoryear{Pojmanski \& Maciejewski}{2004}]{pm04}
Pojmanski G., Maciejewski G., 2004, AcA, 54, 153

\bibitem[\protect\citeauthoryear{Poretti}{1994}]{fp6}
Poretti E., 1994, A\&A, 285, 524

\bibitem[\protect\citeauthoryear{Reegen}{2007}]{reg07} Reegen P., 2007, A\&A, 467,
1353

\bibitem[\protect\citeauthoryear{Samus et~al.}{2002}]{gcvs} Samus 
N. N. et~al., 2002, Astr. Lett., 28, 174

\bibitem[\protect\citeauthoryear{Schmidt et~al.}{2007}]{schetal07}
Schmidt E.~G., Langan S., Rogalla D., Thacker-Lynn~L., 2007, AJ, 133, 665

\bibitem[\protect\citeauthoryear{Simon \& Lee}{1981}]{sile81}
Simon N.~R., Lee A.~S., 1981, ApJ, 248, 291

\bibitem[\protect\citeauthoryear{Spreckley \& Stevens}{2008}]{ss08} Spreckley S.~A., 
Stevens I.~R., 2008, MNRAS, 388, 1239

\bibitem[\protect\citeauthoryear{Strohmeier et~al.}{1963}]{stretal63}
Strohmeier W., Knigge R., Ott H., 1963, Bamberg Ver\"off., V, Nr.~16

\bibitem[\protect\citeauthoryear{Szabados}{1977}]{sza77}
Szabados L., 1977, Commun. Konkoly. Obs. Hung. Acad. Sci., Budapest, No.~70

\bibitem[\protect\citeauthoryear{Szab\'o et~al.}{2007}]{sbb07} Szab\'o R.,
Buchler J.~R., Bartee J., 2007, ApJ, 667, 1150

\bibitem[\protect\citeauthoryear{Szab\'o et~al.}{2010}]{szr10} Szab\'o R., Koll\'ath Z., 
Moln\'ar L. et~al. 2010, MNRAS, 409, 1244

\bibitem[\protect\citeauthoryear{Tammann et~al.}{2003}]{tsr03} Tammann G.~A., 
Sandage A., Reindl B., 2003, A\&A, 404, 423

\bibitem[\protect\citeauthoryear{Voges et~al.}{1999}]{vogetal99}
Voges~W., Aschenbach~B., Boller~Th. et~al., 1999, A\&A, 349, 389

\bibitem[\protect\citeauthoryear{Wachmann}{1976}]{wac76}
Wachmann A.~A., 1976, A\&AS, 23, 249

\bibitem[\protect\citeauthoryear{Wo\'zniak et~al.}{2004}]{wozetal04}
Wo\'zniak P.~R., Vestrand W.~T., Akerlof C.~W. et~al.,
 2004, AJ, 127, 2436

\end{thebibliography}
\end{document}